\documentclass[structabstract]{aa}

\usepackage{epsfig,natbib,url,twoopt}
\usepackage{graphicx}
\usepackage{txfonts}
\usepackage{longtable}
\usepackage{supertabular}
\usepackage{lscape}
\usepackage[breaklinks=true]{hyperref} 
\usepackage[labelfont=footnotesize,textfont=footnotesize]{caption}  

\begin{document}

\title{$Magellan$/MMIRS near-infrared multi-object spectroscopy of nebular emission from star forming galaxies at $2< z < 3$
  \thanks{This paper includes data gathered with the 6.5 meter Magellan Telescopes located at Las Campanas Observatory, Chile}}
\author{L. Guaita\inst{1}
  \and H. Francke\inst{2}
   \and E. Gawiser\inst{3} 
   \and F. E. Bauer\inst{2,}\inst{6}
   \and M. Hayes\inst{4,}\inst{5}
    \and G. {\"O}stlin\inst{1}
 \and N. Padilla\inst{2}}
\offprints{Lucia Guaita, \email{lguai@astro.su.se}}
\institute{Department of Astronomy, Oskar Klein Center, Stockholm University,  AlbaNova, Stockholm SE-10691, Sweden
  \and Pontificia Universidad Cat\'{o}lica de Chile, Departamento de Astronom\'{\i}a y Astrof\'{\i}sica, Casilla 306, Santiago 22, Chile
  \and Department of Physics and Astronomy, Rutgers, The State University of New Jersey, Piscataway, NJ 08854
 \and Universit\'e de Toulouse; UPS-OMP; IRAP; Toulouse, France
  \and CNRS; IRAP; 14, avenue Edouard Belin, F-31400 Toulouse, France
  \and Space Science Institute, 4750 Walnut Street, Suite 205, Boulder, Colorado 80301
}
\date{Accepted date: 14/01/2013}
\abstract
%Context
{}
%Aim
{To investigate the ingredients, which allow star-forming galaxies to present Ly$\alpha$ line in emission, we studied the kinematics and gas phase metallicity of the interstellar medium. }
%Method
{We used multi-object near-infrared spectroscopy with $Magellan$/MMIRS to study nebular emission from $z\simeq2-3$ star-forming galaxies discovered in three MUSYC fields. }
%Results
{We detected emission lines from four active galactic nuclei and 13 high-redshift star-forming galaxies, including H$\alpha$ lines down to a flux of ($4\pm1$)E-17 erg sec$^{-1}$ cm$^{-2}$. This yielded seven new redshifts. The most common emission line detected is [OIII]5007, which is sensitive to metallicity. 
We were able to measure metallicity (Z) for two galaxies and to set upper (lower) limits for another two (two). The metallicity values are consistent with $0.3<Z/Z_{\odot}<1.2$, 12 + log(O/H) $\sim$ 8.2-8.8. Comparing the Ly$\alpha$ central wavelength with the systemic redshift, we find $\Delta v_{Ly\alpha-[OIII]5007}$ = 70-270 km sec$^{-1}$.}
%Conclusions
{High-redshift star-forming galaxies, Ly$\alpha$ emitting (LAE) galaxies, and H$\alpha$ emitters appear to be located in the low mass, high star-formation rate (SFR) region of the SFR versus stellar mass diagram, confirming that they are experiencing burst episodes of star formation, which are building up their stellar mass. 
Their metallicities are consistent with the relation found for $z\leq2.2$ galaxies in the Z versus stellar mass plane. 
The measured $\Delta v_{Ly\alpha-[OIII]5007}$ values imply that outflows of material, driven by star formation, could be present in the $z\sim2-3$ LAEs of our sample. Comparing with the literature, we note that galaxies with lower metallicity than ours are also characterized by similar $\Delta v_{Ly\alpha-[OIII]5007}$ velocity offsets.

Strong F([OIII]5007) is detected in many Ly$\alpha$ emitters.
Therefore, we propose the F(Ly$\alpha$)/F([OIII]5007) flux ratio as a tool for the study of high-redshift galaxies; while influenced by metallicity, ionization, and Ly$\alpha$ radiative transfer in the ISM, it may be possible to calibrate this ratio to primarily trace one of these effects. }


\titlerunning{NIR spectroscopy of star-forming galaxies}
\authorrunning{L. Guaita}
\maketitle

%%%%%%%%%%%%%%%%%%%%%%%%%%%%%%%%%%%%%%%%%%%%%%%%%%%%%%%%%%%%%%%%%%%%%%%%%%%%
\section{Introduction}     \label{sec:introduction}
%%%%%%%%%%%%%%%%%%%%%%%%%%%%%%%%%%%%%%%%%%%%%%%%%%%%%%%%%%%%%%%%%%%%%%%%%%%%

Characterizing star formation at high redshift is critical for the investigation and understanding of galaxy formation and evolution. This task is often difficult, however, since distant star-forming galaxies can be quite faint and many of their characteristic diagnostic features redshifted into different observational regimes. The identification of high-redshift objects during specific evolutionary stages amidst foreground sources generally falls into two categories, either photometric methods which sample portions of the integrated spectral energy distribution (SED) or narrow-band imaging which select specific emission lines at specific redshifts. 

Photometric methods (e.g., the Lyman break galaxy, or ‘LBG’, technique, Steidel et al. 1996; the Lyman alpha decrement, or ‘BX’ technique, Steidel et al. 2004) have been extensively used to identify large samples of high-redshift galaxies, and even characterize their physical properties via comparisons to theoretical SEDs, generated by various initial mass functions, star-formation histories, stellar masses, dust reddening and interstellar medium (ISM) metallicities \citep[e.g.,][]{Shapley:2003,Pentericci:2007}. The advantage of the photometric techniques is that faint sources, occupying wide fields-of-view and wide redshift slices, can be detected, while the disadvantage is the lack of redshift confirmation.

A more efficient way to detect galaxies, at least in a narrow redshift range, is by using narrow-band filters covering a specific line wavelength at a specific redshift, such as Ly$\alpha$ \citep[e.g.,][]{CHu1998}. In principle Ly$\alpha$ could be observed up to $z = 10$ from the ground. Over the past decade, astronomers have built large samples of Ly$\alpha$ emitters (LAEs) at $z>2$ \citep[e.g.,][]{Rhoads:2000,Gronwall:2007,Ouchi:2008,Nilsson:2009}. Narrow-band techniques have been used to detect H$\alpha$, [OII]3727 and [OIII]5007 line emitters at various redshifts \citep[e.g.,][]{Thompson1994,Hayes2010b,Hatch2011,Shim2011}. Such surveys, however, often require spectroscopic follow-up to weed out lower redshift emission-line galaxies which contaminate the samples. New multi-narrow-band surveys, with narrow-band filters covering Ly$\alpha$, [OII]3727 and H$\alpha$+[NII] at $z=2.2$ \citep[e.g.,][]{Nakajima2012b,Lee2012}, now have the advantage to reduce almost completely the contamination from low-redshift objects and also to calculate emission-line ratios, sensitive to galaxy metallicity.

In the context of the MUSYC survey \citep{Gawiser:2006a}, \footnote{http://physics.rutgers.edu/$\sim$gawiser/MUSYC} we amassed two samples of LAEs at $z\simeq3.1$ and $z\simeq2.1$. Their physical properties were studied by \citet{Gronwall:2007}, \citet{Gawiser:2007}, \citet{Lai:2008}, \citet{Guaita2010}, \citet{Guaita2011}, \citet{Acquaviva2012}, and \citet{Bond2012}, which found LAEs to be in the lower mass range of the mass distribution of star-forming galaxies, observed during active phases of star formation, and characterized by low dust content and compact morphology. These are the ideal samples to constrain any evolution (or lack thereof) between redshifts of $z\simeq3.1$ and $z\simeq2.1$. 
\citet{Ciardullo2012} found $z\simeq3.1$ LAEs with luminosities brighter than L(Ly$\alpha)>10^{42.4}$ erg sec$^{-1}$ had almost twice the number density of $z\simeq2.1$ LAEs in the Extended $Chandra$ Deep Field South. Also, the number of galaxies with EW(Ly$\alpha)  > 100$ {\AA} at $z\simeq3.1$ was almost twice the number at $z\simeq2.1$. 
While $z\sim3$ LAEs were observed to be consistent with zero dust content \citep{Gawiser:2007,Nilsson:2009,Ono2010}, the SED fitting of LAEs at $z\sim2$ showed a moderate amount of dust \citep{Nilsson2010,Hayes2010,Guaita2011}. 

The SED analyzes also showed that $z\simeq3.1$ LAEs are low mass ($10^8-10^9$ M$_{\odot}$), young galaxies (20 Myr) in active phases of star formation \citep{Gawiser:2007,Lai:2008}, while $z\simeq2.1$ LAEs are instead characterized by more variety. 
For instance, some $z\simeq2.1$ LAEs have `ages since star formation began' that are compatible with a few hundreds of Myr and the sub-sample of more massive galaxies ($10^{9.5}$ M$_{\odot}$) are characterized by redder colors (e.g., Gu2011). This was interpreted as an indication that there are a wider variety of ISM mechanisms, due to the spatial and velocity distribution of dust and neutral Hydrogen at lower redshift, which could be responsible for the escape of Ly$\alpha$ photons \citep[with delayed episodes of star formation among the proposals;][]{shimizu2010}.
A recent scenario was described in \citet{Acquaviva2012}, who used a Monte Carlo Markov chain code \citep[GalMC,][]{Acq2011} to derive physical parameters from the SEDs of our $z\simeq3.1$ and $z\simeq2.1$ LAE samples. They proposed that higher-redshift LAEs were on average more metal-poor, even if older, than lower redshift ones. The investigation of intrinsic properties and direct metalllicity estimations for individual galaxy could be very useful to confirm this hypothesis.
 Our current work here looks to address two main aspects of this scheme via estimates of the gas phase metallicity and an investigation of ISM kinematics. Star-formation activity and supernova explosions could be responsible for pushing material out of the galaxies in the form of outflows, which consequently leave imprints in the profile of Ly$\alpha$ emission \citep{V:2006}.

The estimation of metal line abundances is important to increase the statistics of metallicity estimations in star-forming galaxies (SFGs) at high redshift. While Balmer emission lines can be modeled by stellar synthesis codes \citep{SB2011}, a deeper investigation of the [OIII] emission relative to the Hydrogen lines is useful to allow better SED modeling of optical emission lines. 
Notably, strong [OIII]5007 emission can comprise a significant fraction of the $K$ band magnitude at $z\sim3$ \citep{McLin2011}, making its characterization critical for the correct interpretation of broad-band photometric properties. Moreover, \citet{McLin2011} and \citet{Fin2011} recently showed that significant velocities offsets between Ly$\alpha$ and systemic redshift as measured by [OIII] can be present in LAEs, which are clearly important for an interpretation of the Ly$\alpha$ emission.

This pilot project focuses on obtaining rest-frame optical spectra for a sizeable sample of high-redshift galaxies via near-infra-red multi-object spectrographs (NIR-MOS). Only a few NIR-MOS instruments have been available to date. NIR spectra have been obtained by using the multi-object infra-red camera and spectrograph (MOIRCS) at the Subaru telescopes to observe cluster stars \citep[e.g.,][]{Scholz2009},  medium redshift massive galaxies \citep[e.g.,][]{Onodera2010,Hayashi2011}, BzK galaxies with $K_{AB}\leq22$ \citep{Yoshikawa2010}, H$\alpha$ emitters \citep{Tanaka2011}, and quasars \citep{DelMoro2009}. Recently the MMT and $Magellan$ Infrared Spectrograph (MMIRS) was used to obtain NIR spectra of LAEs at $z\simeq2.2$ \citep{Hashimoto2012} with relatively bright optical and NIR magnitudes (at least compared to those observed here).

We present here a new NIR spectroscopic survey, with five main aims: \\
1) to demonstrate the capabilities for studying faint emission-line galaxies with MMIRS on the $Magellan$ Clay 6.5m telescope;\\
2) to confirm redshifts of photometrically-selected SFGs; \\
3) to measure rest-frame optical emission line ratios and estimate metallicities of galaxies undergoing active phases of star formation;\\
4) to place SFGs, previously detected via Ly$\alpha$ or H$\alpha$ emission lines, in an evolutionary sequence and investigate similarities and differences with respect to continuum-selected SFGs in the same redshift range;\\
5) to probe the gas kinematics of LAEs in order to study channels of the escape of Ly$\alpha$ photons.

We adopt AB magnitudes throughout, following the Two Micron All Sky Survey (2MASS) zero-point magnitudes, $K_{AB} = K_{Vega} + 1.87$.

When we refer to the rest-frame wavelength of emission lines, we use vacuum-wavelengths of $\lambda_{Ly\alpha}=1215.67$ {\AA}, $\lambda_{[OIII]5007}=5008.239$ {\AA}, $\lambda_{H\alpha}=6564.614$ {\AA}. 
We denote the stellar reddening as E(B-V) and the nebular reddening as Eg(B-V). We assume a $\Lambda$CDM cosmology consistent with the WMAP 5-year results \citep[][their Table 2]{Dunkley:2009}, adopting mean parameters of $\Omega_m=0.26$, $\Omega_{\Lambda}=0.74$, H$_0=70$ km~s$^{-1}$~Mpc$^{-1}$, $\sigma_8=0.8$.

%%%%%%%%%%%%%%%%%%%%%%%%%%%%%%%%%%%%%%%%%%%%%%%%%%%%%%%%%%%%%%%%%%%%%%%%%%%%
\section{Observations and sample definition}    \label{sec:obs}
%%%%%%%%%%%%%%%%%%%%%%%%%%%%%%%%%%%%%%%%%%%%%%%%%%%%%%%%%%%%%%%%%%%%%%%%%%%%

MMIRS is a near-IR imager and multi-object spectrograph with a $4'\times6.9'$ field-of-view, \footnote{http://www.cfa.harvard.edu/mmti/mmirs/Calibration/waverange\_HK\_1.png} operating on the $Magellan$ Clay telescope since the end of 2010. Mask design followed the MMT mask preparation procedure. \footnote{http://hopper.si.edu/wiki/mmti/MMTI/MMIRS/ObsManual/\\MMIRS+Mask+Making}
The objects studied in this paper come from two two-night observing runs in late 2010 and mid 2011 (see Table \ref{tab:log}). Weather conditions and instrument performance were more favourable in the first run. Two entire nights were allocated in Nov, while only half nights in May, 19th and 20th. The half nights allocated to the second run resulted in airmasses that were generally higher than the first run, reaching up to airmass 2.0 on May 20th. 

We obtained spectra in three fields belonging to the MUSYC survey \citep[][hereafter Ga2006]{Gawiser:2006a}: two masks in the ECDF-S (Extended $Chandra$ Deep Field South), one mask in the EHDF-S (Extended $Hubble$ Deep Field South), and one mask in the SDSS-1030 region. 

In Table \ref{tab:det} we show the number of detected sources versus the targeted ones. 
In Table \ref{tab:list} we list the 17 sources from which we detect emission lines. The names designed in column 1 will be used hereafter. When filling the masks we chose three main categories of sources:

1) Narrow-band selected LAEs from Gronwall et al. (2007; hereafter Gr2007) at $z\simeq3.1$ and in Guaita et al (2010; hereafter Gu2010) at $z\simeq2.1$. Also, we considered Ly$\alpha$ and/or H$\alpha$ emitters detected by Hayes et al. (2010) in their double blind (DB) survey. The detected sources originally belonging to this category and studied here are LAE27, z3LAE2, z2LAE2, z2LAE3, DB8, DB12, and DB22 (Table \ref{tab:list} notation).

2) SFGs as defined in Ga2006 and \citet{Guaita2011}, and studied in \citet{Berry2012}, selected using the broad-band color drop-out technique at $z\sim3$ (LBGs) and Ly$\alpha$ decrement at $z\sim2$ (`BXs'). The detected sources originally belonging to this category and studied here are BX1, BX4, BX10, BX14, BX16, and LBG11 (Table \ref{tab:list} notation).  Our spectroscopy revealed the presence of Ly$\alpha$ emission with equivalent widths larger than 20 {\AA} for BX1 and BX10, confirming them as LAEs. Also included in this category were BzK galaxies either from \citet{Blanc:2008} with $K_{AB}<22$ or, in the case of Bzk25, selected from deeper VLT infrared spectrometer and array camera (ISAAC) imaging with fainter continua (Mark Dickinson and Jeyhan Kartaltepe, private communication). The latter was not expected a priori to present any continuum in the NIR. 

3) Active galactic nuclei (AGNs) expected to lie in the range $2<z<3$ and show strong rest-frame optical lines, selected from optical \citep{Francke:2008}, radio \citep{Miller2008}, or X-ray catalog \citep{Xue2011}. The AGNs originally belonging to this category and studied here are AGN5 and AGN26.

We designed the MMIRS masks giving first priority to Ly$\alpha$/H$\alpha$ emitting galaxies observed at $z\simeq2$ and $z\simeq3$. 

Galaxies selected to be either Ly$\alpha$ and/or H$\alpha$ emitters were expected have faint continua and only display a few emission lines in the NIR spectra like [OIII] and H$\alpha$ due to star formation. NIR spectra of DB sources, where H$\alpha$ fluxes were measured photometrically, are very important for setting upper limits on H$\alpha$ detection. We filled in the remaining space on the masks using SFGs and AGNs with or without firmly known redshifts but suspected to lie at $z\geq1.5$. We also included low-redshift galaxies, which were expected to yield strong continuum and thus be useful tests of MMIRS' capability, as well as slit stars for mask alignment phase and flux calibration. In each of the four masks we included 3-4 square boxes with alignment stars, although during the second run we used three additional stars for flux calibration and additional alignment checks. 

Our $z\simeq2.1$ and $z\simeq3.1$ LAE samples had a median Ly$\alpha$ flux of $4.0 \times10^{-17}$ erg sec$^{-1}$ cm$^{-2}$ (Gr2007, Gu2010). 
The Ly$\alpha$/H$\alpha$ ratio can be significantly reduced from the theoretical value of 8.7 for case B recombination \citep{Oster1989} even by a small amount of dust reddening. To produce a reasonable estimate of the reddening expected, we assumed a Calzetti law with E(B-V)=0.04, which is typical of our sample LAEs (Ga2007, Gu2011), and that Ly$\alpha$ photons experience the enhanced dust attenuation of nebular gas but no further amplification of dust column from radiative transfer.  This predicts a Ly$\alpha$/H$\alpha$ ratio of $\sim$3, implying a median H$\alpha$ flux for our sample of $1.3\times10^{-17}$ erg sec$^{-1}$ cm$^{-2}$. Scaling from the predicted S/N per resolution element for the current HK grism (MMIRS S/N calculator, \footnote{http://www.cfa.harvard.edu/mmti/mmirs/instrstats.html}) we anticipated a S/N of 6 in 4 hours for the lines predicted in our typical LAEs. We obtained total exposure times of 4.8 and 4.7 hours for our two ECDF-S masks, and 3.2 and 2.7 hours for our SDSS and EHDF-S masks, respectively. Spectra were taken through the $HK$ grism and $H+K$ filter covering 1.25-2.45 $\mu m$ wavelength range, through a slit with width equal to 0.7'' (0.5'' for a few sources in ECDF-S masks). The filter transmission has the same throughput  of 70\% at 16000 and 21000 {\AA}. At 19000 {\AA} the throughput has its maximum, but the atmospheric absorption is maximum at this wavelength as well. 

During the Nov run we chose a dithering pattern of 0, 3, -2, 2, -3 arcsec along the slit length to perform a better sky subtraction. 
However, the shift of pair subtracted images could cause an overlap of close-by slits (Sect. \ref{sec:red}). Also, a 3'' dither caused alignment stars to move outside of their boxes in a few cases. Therefore, we adopted a 0, 2, -2 pattern in the May run.  We re-aligned the masks frequently during the nights, typically after a series of 5-10 science frames.
We observed telluric standard stars\footnote{http://hopper.si.edu/wiki/mmti/MMTI/MMIRS/MMIRS+Pipeline/\\Telluric+Correction} 2-3 times per mask per night, placing the telluric in 5 slits across the mask to achieve full spectroscopic coverage.

MMIRS is characterized by a 0.2012''/pixel scale. It corresponds to a 3.5 (2.5) pixel resolution element for a 0.7'' (0.5'') slit width. For a sampling of 7 {\AA}/pixel, this means a line resolution of 25 (18) {\AA} and a resolution R=800 (1100) at $\lambda=20000$ {\AA}.

%%%%%%%%%%%%%%%%%%%%%%%%%%%%%%%%%%%%%%%%%%%%%%%%%%%%%%%%%%%%%%%%%%%%%%%%%%%%
\section{Reduction}    \label{sec:red}
%%%%%%%%%%%%%%%%%%%%%%%%%%%%%%%%%%%%%%%%%%%%%%%%%%%%%%%%%%%%%%%%%%%%%%%%%%%%

A full account of our reduction procedure is reported in the Appendix \ref{sec:appendix}, while here we describe only the main details.

We used the $mmfixen$ code provided by the MMIRS instrument scientific team\footnote{http://hopper.si.edu/wiki/mmti/MMTI/MMIRS/\\MMIRS+Pipeline} to collapse all the information enclosed in the multi-extension files. The output became a new multi-extension file, in which the 1st extension was the science frame normalized to counts/sec.

To calculate the wavelength calibration map we used the COSMOS{\footnote{http://hopper.si.edu/wiki/mmti/MMTI/MMIRS/\\MMIRS+Pipeline/MMIRS+COSMOS}}
public software package, which was originally designed to reduce optical multi-object spectra obtained with the $Magellan$ IMACS and LDSS instruments and updated to work with NIR MMIRS data. Positional files (*.SMF) were generated for each mask by the MMIRS team (and mmirs2smf.py code) to make MMIRS mask frames compatible with COSMOS.  We obtained a rough wavelength map using COSMOS  $align$-$mask$ and $adjust$-$offset$ codes and then a final solution running COSMOS $adjust$-$map$. 

The most prominent features in our NIR spectra were the atmospheric sky lines. They were used as a reference for the wavelength calibration, but afterwards they were eliminated from the science frames using an $ABBA$ source offset procedure. 
We first subtracted two frames, A and B, adjacent in time to remove the bulk of the constant and variable IR backgrounds, obtaining background-subtracted (A-B) or (B-A) frames. Depending on the sky variation at a specific moment of the night and airmass, in the A-B and B-A frames sky lines are reduced by about 50-80\%. 
We then applied offsets to align the dispersed source spectra from the A and B frames and summed the signals of aligned frames of the same mask for the entire run. The shift was determined based on the input dithering pattern. The remaining sky line residuals were due to rapid temporal sky line variations and variations along the slit for individual lines. These variations dominate the background noise. 

We then applied the previously calculated wavelength map to the completely reduced frames. This was performed using the routine COSMOS $extract$-$2dspec$, which applies the wavelength map and and splits a multi-slit frame into individual slit 2D spectra. We used the IRAF $twodspec.apextract.apall$ task to extract 1D spectra. 

Giant A0 stars (without strong features in their NIR spectra) were used as flux standards to generate the sensitivity function needed for flux-calibrating science frames, as well as to correct telluric absorption features.

We chose extraction $apall$ apertures of $\pm$3 ($\pm$5) pixels for extracting 1D standard and science source spectra, respectively. The fractional missing flux due to the wings of the PSF falling outside the extraction aperture (not to be confused with slit losses) is estimated to be less than 10\% in both the runs.

In the case of the Nov run, we compared our flux calibration for emission lines from AGN5 to those reported in the literature by \citet{VanDokkum2005}; see Sect. \ref{sec:performance} for more details. In the case of the May run, we compared the obtained sensitivity functions for our slit stars with their literature magnitudes, finding good agreement.

%%%%%%%%%%%%%%%%%%%%%%%%%%%%%%%%%%%%%%%%%%%%%%%%%%%%%%%%%%%%%%%%%%%%%%%%%%%%
\section{Emission line fitting procedure}    \label{sec:fit}
%%%%%%%%%%%%%%%%%%%%%%%%%%%%%%%%%%%%%%%%%%%%%%%%%%%%%%%%%%%%%%%%%%%%%%%%%%%%

After extracting 1D spectra, we fitted emission lines and continua with the python scipy $optimized.leastsq$ function. 
The parameters of the fit were central wavelength ($\lambda_0$), dispersion ($\sigma_{gauss}$), and amplitude (A$_{gauss}$) of a Gaussian curve (f$_{\lambda}$ = (A$_{gauss}$/$\sqrt{2\pi}$) exp[-($\lambda-\lambda_0)^2/2\sigma_{gauss}^2]$) for emission lines and intercept of a flat continuum.

To easily identify the first-guess fitting parameters we smoothed all raw spectra by a $\sim$30 {\AA} box. We performed a linear fit to the smoothed continua at $\sim$16000 {\AA} and $\sim$21000 {\AA}, excluding emission line wavelength regions from the continuum fit. We fitted the smoothed-spectrum continuum-subtracted H$\alpha$ and [OIII]5007 emission lines with single Gaussians to measure $\lambda_0$ (hence the source's redshift) and the first-guess $\sigma_{gauss}$ and A$_{gauss}$. 

We performed the fit a second time to obtain best-fit $\sigma_{gauss}$ ($\sigma_{gauss,best-fit}$) values for three Gaussians fitted to H$\alpha$+[NII]doublet simultaneously, fixing the redshift inferred from $\lambda_0$ in the previous step and varying $\sigma_{gauss}$ and A$_{gauss}$ around the first-guess values. We fixed the ratio of the [NII] lines to 1:3 \citep{Oster1989}. Even if [NII] is not clearly seen in the 2D spectrum, this procedure was important to estimate the contribution of [NII] blended to H$\alpha$. \citet{Atek2011} found that the [NII]doublet contribution to H$\alpha$ was on average 8\% in their sample. We similarly fitted the [OIII]doublet + H$\beta$ with three Gaussians, where we fixed the [OIII]doublet ratio to 1:2.98 \citep{SZ2000}. We excluded pixels strongly contaminated by sky-line residuals from the fit.

After obtaining the best-fit $\lambda_0$ and $\sigma_{gauss,best-fit}$ for the six emission lines in the smoothed spectrum, we fixed these values and searched for the best-fit A$_{gauss}$ (A$_{gauss,best-fit}$) in the raw spectrum. 
From this last amplitude estimation we calculated the integrated line flux (F(line)=A$_{gauss,best-fit} \times \sigma_{gauss,best-fit} \times \sqrt{2 \pi}$). We estimated the 1$\sigma$ uncertainty on the $\lambda_0$, $\sigma_{gauss,best-fit}$, and A$_{gauss,best-fit}$ values assuming a single parameter of interest, calculating the diagonal elements of the fit covariance matrix.

The [OIII]doublet was a robust feature to find because it is composed by two separate emission lines with a known flux ratio. If we saw [OIII], but not H$\alpha$, we assumed an H$\alpha$ upper limit of twice the continuum root-mean-square (rms) error at the wavelength of H$\alpha$ multiplied by the MMIRS resolution of 25 {\AA}. The same method was applied in the case we detected H$\alpha$ but not [OIII].

%%%%%%%%%%%%%%%%%%%
\subsection{Uncertainties of the emission line parameters}
\label{error}
%%%%%%%%%%%%%%%%%%%

To estimate line integrated flux uncertainty, we used the parameter uncertainty derived from the best fit covariance matrix (Sect. \ref{sec:fit}). However, additional factors could increase the total error budget.

We performed a few tests to verify the goodness of the reduction and calibration processes. We calculated the observation point spread function (PSF) by measuring the median full width half maximum (FWHM) of the telluric standard spectral profile.  Because telluric standards were observed in 5 slits along the mask spatial axis, we could estimate the average PSF variation along the mask. In addition to this, we could estimate the possible variations of the FWHM spectral profile with time in the series of 5 sub-sequent exposures. This could be related to the alignment, which might have varied with time. But in a series of five 1 sec-long exposures, $\Delta FWHM$ was just 0.1''. We did the same test considering the spectral profiles of stars in the science masks as well. The FWHM and the $\Delta FWHM$ are comparable to the ones measured for telluric stars. In the Nov run, for example, FWHM=0.8''$\pm$0.2'' for telluric and science frames. This means that our flux calibration could compensate for slit losses, in the case extraction aperture and slit width were the same for telluric and science spectra. The fraction of the flux in the wings of the spectral profile ending outside the extraction aperture should be less than 10\%. 

Flux calibration uncertainties are estimated to be less than 10\% in the Nov masks and up to 30\% in the May masks. This was produced by the standard deviation of the flux among standard spectra taken from the same slit along the run, due to changes in weather conditions.

Sky line residuals could also affect the integrated flux estimation. We avoided significant pixels with residual emission from the sky in the line fitting. In the case of [OIII], the presence of a doublet and a fixed ratio 2.98:1 compensated for the potential increase/decrease of flux due to sky line residuals.

Depending on the night seeing/slit width, sky emission lines can be broadened, producing an uncertainty in the wavelength calibration and so in the observed line center as well. 
The wavelength map needed to wavelength calibrate our spectra was successfully generated through the COSMOS package, but an uncertainty of 10 {\AA} is the upper limit in the wavelength uncertainty due to broadened sky lines. To estimate this value, we extracted sky spectra from random slits of a few science frames, in which the $ABBA$ procedure was not applied yet. We also average combined all these science frames the same way the final reduced ones were combined and extracted the sky spectrum from the combination. Then we compared the emission line wavelengths with the tabulated ones.

All these uncertainties are included in central wavelength and integrated line flux total error budgets, reported in the following figures.

%%%%%%%%%%%%%%%%%%%%%%%%%%%%%%%%%%%%%%%%%%%%%%%%%%%%%%%%%%%%%%%%%%%%%%%%%%%%
\section{Sources detected in MMIRS spectra}    \label{sec:performance}
%%%%%%%%%%%%%%%%%%%%%%%%%%%%%%%%%%%%%%%%%%%%%%%%%%%%%%%%%%%%%%%%%%%%%%%%%%%%

In Table \ref{tab:list} we list the 17 sources, for which we detected emission lines (H$\alpha$ and [OIII]5007 have signal-to-noise ratio, S/N $\geq3$). 
AGNs, LAEs, and DB galaxies had the highest fraction of successful detections. The main reason for the low fraction was that the sources were generally faint in $K$ band and that in some cases a particular emission line was overlapped with a strong sky line residual.

\begin{figure*} 
\centering
\includegraphics[width=150mm]{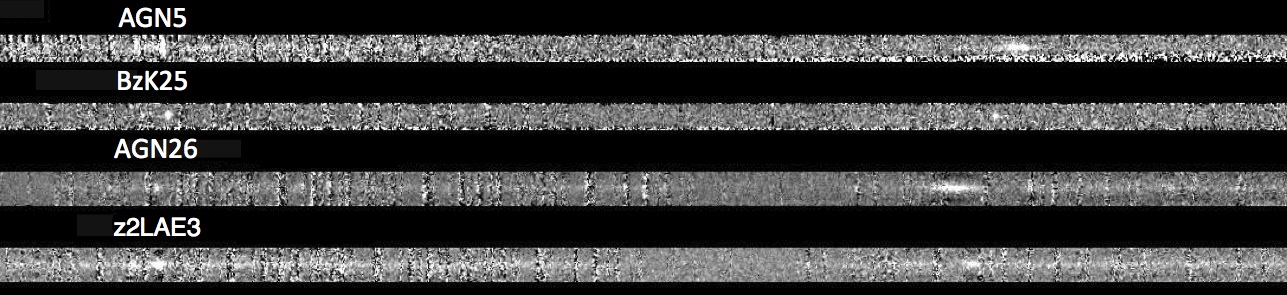}
\caption{2D spectra of AGNs detected in the 4 masks, all characterized by $K_{AB}>20$. The spectra are aligned in wavelength. Poisson noise due to sky emission line residuals can be seen as black vertical stripes.}
\label{AGN2d}
\end{figure*}

\begin{figure*} 
\centering
\includegraphics[width=120mm]{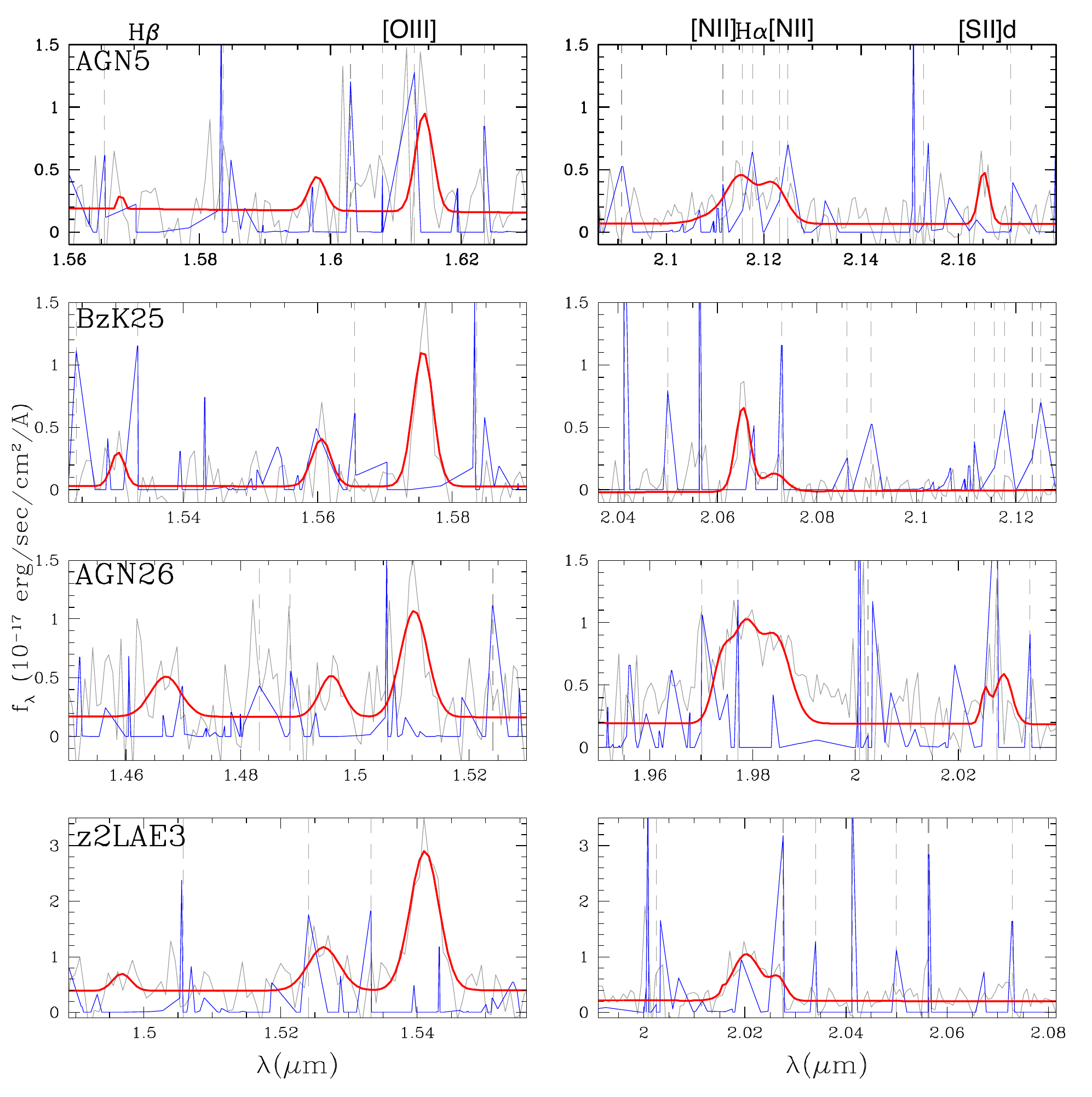} 
\caption{Observed 1D spectra of AGNs in grey and best fit in red. From the top AGN5, BzK25, AGN26, z2LAE3. Vertical dashed lines represent the wavelengths of sky line residuals. The blue curve is the sky spectrum by \citet{rou2000}, which indicates the relative intensity of the OH emission lines. It has been scaled according to the panel axis to be visible. [OIII]doublet (H$\alpha$) wavelength ranges are presented in the left (right) panels.}
\label{ELagn}
\end{figure*}

Several of the targetted AGNs exhibited strong, broad emission lines clearly seen in their 2D spectra (Fig. \ref{AGN2d}). Therefore, these sources were good candidates to test the sensitivity and capability of MMIRS, as described below. 
 In Fig. \ref{ELagn} we show the AGN spectra in black and we over-plotted the best fit Gaussian curve in the [OIII] and H$\alpha$ wavelength region as red curves. The blue spectrum is the sky spectrum from \citet{rou2000}. Fig. \ref{agn} shows the line flux ratio diagram from \citet{Kewley2001} where SFGs appear to be separated from AGNs (upper left side in each panel). We used this diagram \citep[originally proposed by][]{Baldwin1981} to either identify and/or confirm AGNs among our sources.

%%%%%%%%%%%%%%%%%%%
\subsection{Active galactic nuclei}
%%%%%%%%%%%%%%%%%%%

{\bf{AGN5}} was previously studied in \citep[][hereafter vD05]{VanDokkum2005}, using Gemini/GNIRS. They used the GNIRS short camera with 32 lines mm$^{-1}$ and 0.68'' wide, 6.2'' long slit. Their total exposure time was $\sim$5500 sec, with a seeing of $\sim0.7$''.  We used a 0.5'' slit in the $H+K$ grism configuration for a total exposure time of 17280 sec (4.8 hours). The estimated total spectroscopic throughput is $\sim30$\% in $K$ and $\sim35$\% in $H$. 
This afforded us the chance to cross-check the performance of MMIRS against high-quality Gemini/GNIRS data. 
We mentioned before that our flux calibration was able to compensate for slit losses. By definition it also accounted for telescope aperture and system efficiency. However, for this object we needed to take into account an additional slit loss due to a narrower slit of 0.5'' during a night with 0.8'' seeing. This implied that we needed to increase the integrated flux by about 20\% to compare with van Dokkum's results. As described in the previous section, we fitted the continuum-subtracted emission lines with Gaussian curves to obtain the integrated flux and its uncertainty, after correcting for slit losses. Sky emission line residuals at 16128 {\AA} were excluded from the fit of the [OIII] doublet and we obtained F([OIII]5007) $=32.5\pm6.7$ E-17 erg sec$^{-1}$ cm$^{2}$, F([OIII]4959) $=10.9\pm2.3$ E-17 erg sec$^{-1}$ cm$^{2}$. Faint sky lines lying in the middle of the H$\alpha$+[NII] complex at 21116 and 21156 {\AA} might still affect the results. We measured F(H$\alpha$) $=28.1\pm7.9$ E-17 erg sec$^{-1}$ cm$^{2}$. 
The continuum rms in the [OIII] and H$\alpha$ wavelength ranges are not significantly different.

Considering a PSF variation ($\Delta FWHM$) of the order of 0.2'' during the run and a flux calibration uncertainty of 10\%, the MMIRS line flux measurements are consistent with the ones obtained using GNIRS by vD05 (F([OIII]4959)=(10.0$\pm$1.0)E-17 erg sec$^{-1}$ cm$^{2}$, F([OIII]5007)=(24.0$\pm$2.4)E-17 erg sec$^{-1}$ cm$^{2}$, and F(Ha)=(24.0$\pm$2.4)E-17 erg sec$^{-1}$ cm$^{2}$). 

\begin{figure}[h!]
\centering
\includegraphics[width=80mm]{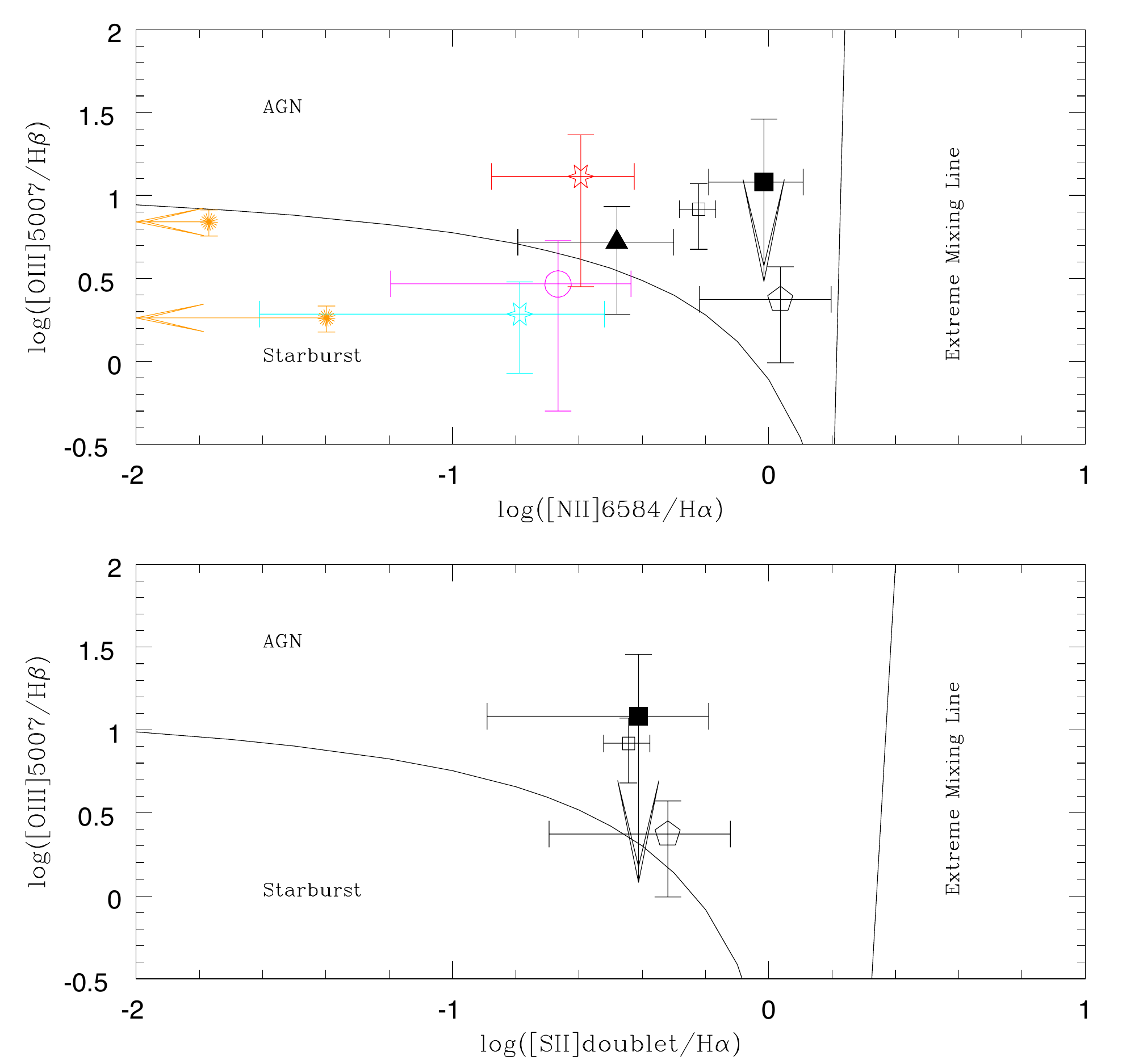}  
\caption{Line ratio diagrams as presented in \citet{Kewley2001} (solid lines). The solid lines divide $AGNs$ from $Starburst$ galaxies. Symbols indicate AGNs found in our survey as explained in the text, the filled (open) squares are our \citep{VanDokkum2005} ratio estimations for AGN5, the black triangle (pentagon) represents Bzk25 (AGN26), the red star z2LAE3. SFGs located in the region typical of $Starbursts$ are presented for comparison: DB12 as a 
magenta circle, BX16 as a cyan star, and the two LAEs from \citet{Fin2011} as orange dots. }
\label{agn}
\end{figure}

The squares in the upper and lower panels of Fig. \ref{agn} represent our estimations of line ratios for AGN5 (filled) and previous van Dokkum's (open). Even if H$\beta$ has very low S/N in our spectrum and so [OIII]/H$\beta$ ratio is consistent with an upper limit, AGN5 appear to be located in the $AGN$ region of the diagram. 

{\bf{BzK25}} was originally selected as a BzK galaxy. A public optical spectrum was available from the GOODS VIMOS survey 
\citep[GOODS\_LRb\_002\_1\_q4\_79\_1.spec, Quality A optical redshift of 2.1449,][]{Balestra2010}. \footnote{http://www.eso.org/sci/activities/garching/projects/goods/\\spectroscopy.html}

This showed CIII]1909 and a weak HeII in emission. This spectrum might indicate it was an AGN, although CIV and NV were not formally detected (we note there is a faint line at the location of CIV which appears to have a P-Cyg profile; such features can be observed for SFGs). The MMIRS spectrum did not show any continuum, but showed [OIII] and H$\alpha$ emission lines (S/N([OIII])=7, S/N(H$\alpha)$=10). The filled triangle in Fig. \ref{agn} represents this source and supports the idea it is an AGN.
Looking at Fig. \ref{ELagn}  we can see sky lines are not significantly affecting H$\alpha$ or [OIII]5007. 

{\bf{AGN26}} is represented by the open pentagon in Fig \ref{agn}. The line ratio diagram, we chose to show here, indicates it is an AGN. The [OIII] doublet is not affected by sky emission lines, but sky line residuals could affect the [SII] integrated flux. The red star corresponds to {\bf{z2LAE3}}. Even if it was initially selected as LAE through narrow-band technique in EHDF-S, it appeared to be an AGN based on Fig. \ref{agn} line ratios. It is the brightest source of the sample, with L(Ly$\alpha$) was equal to 10$^{43.5}$ erg sec$^{-1}$ (F(Ly$\alpha$) $\simeq$ 1.0E-15 erg sec$^{-1}$ cm$^{2}$  and EW$(Ly\alpha)_{rest-frame}$ = 74 {\AA}) and it was the brightest source of the LAE sample in EHDF-S. 

For comparison, we show BX16 and DB12, which fall in the $Starburst$ region of the diagram, as well as LAEs from \citet{Fin2011}. However, we note that the most massive and most Ly$\alpha$ luminous galaxies are generally found near the border of $AGN$ and $Starburst$ regions.

\subsection{Star-forming galaxies} 
\label{sec:slits}

We adopted the same procedure described in Sect. \ref{sec:fit} and \ref{sec:performance} for fitting continuum and emission lines of all the sources listed in Table \ref{tab:list} and showed in Figs. \ref{ELLAE} and \ref{ELother}. 
The SFGs detected in our MMIRS spectra showed an average $K_{AB}$ magnitude of 22.3. We were not able to detect either emission lines nor continua from BzK galaxies fainter than $K_{AB}=23$ or photometrically selected LAEs with L(Ly$\alpha$) fainter than 10$^{42.94}$ erg sec$^{-1}$.
The most prominent emission line observed is [OIII]5007 (typical S/N $=3-5$), which is very intense in SFGs \citep{SB2011} and easy to identify thanks to its doublet. In Table \ref{tab:ELflux} we report line fluxes, redshift, observed-frame equivalent width, and luminosities. This information was used to characterize the properties of SFGs in our sample.

\subsubsection{Physical properties}
\label{SEDpro}

In the past six years our sample of SFGs at $2<z<3$ was studied via photometry and optical spectroscopy. 
In Gu2011 we fitted their stacked SED and showed that stellar mass and reddening were the only parameters we were able to constrain with high significance. We found that the median stacked SEDs were able to reproduce the median stellar E(B-V) and mass values of a sample, we defined as the typical ones. We accounted for the spread in the sample population by using the boot-strapping method to evaluate the uncertainties. It was also pointed out that stacking increased the signal-to-noise of ground-based photometric SEDs and revealed typical properties of our LAE samples. We needed LAE sub-samples (based on rest-frame UV magnitude, color, Ly$\alpha$ equivalent width, and brightness at observed-frame 3.6 $\mu$m separations) to identify unusual galaxy properties different from the typical ones.

\begin{figure*} 
\centering
\includegraphics[width=150mm]{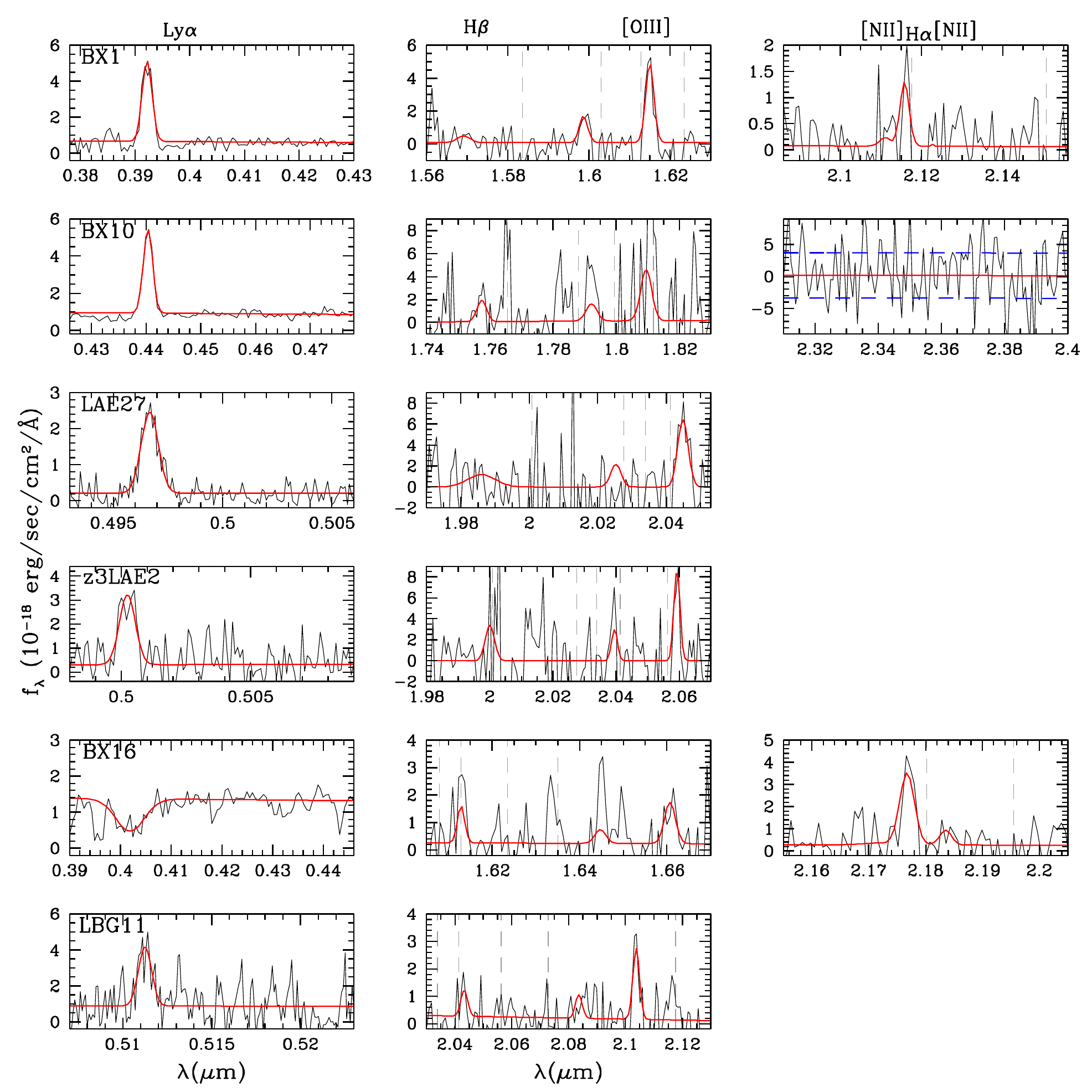}
\caption{Observed-frame 1D spectra (black curve) of the sources showing either Ly$\alpha$ in emission or in absorption in rest-frame UV spectra. In red we show the best fit model spectra. 
From the top BX1, BX10, LAE27, z3LAE2, BX16, LBG11. The first, second and third panel columns represent Ly$\alpha$ (from MUSYC and GOODS-VIMOS surveys), [OIII]doublet and H$\alpha$ wavelength ranges respectively. 
Dashed vertical lines indicate the wavelength of sky emission lines, which could leave residuals in the science spectra.
 Horizontal dashed blue lines indicate the continuum rms in the wavelength range where we did not see an expected emission line. As LAE27, z3LAE2, and LBG11 are at $z\sim3$, H$\alpha$ line would fall outside the MMIRS coverage. The rest-frame optical emission lines are unresolved.}  
\label{ELLAE}
\end{figure*}

\begin{figure*} 
\centering
\includegraphics[width=85mm]{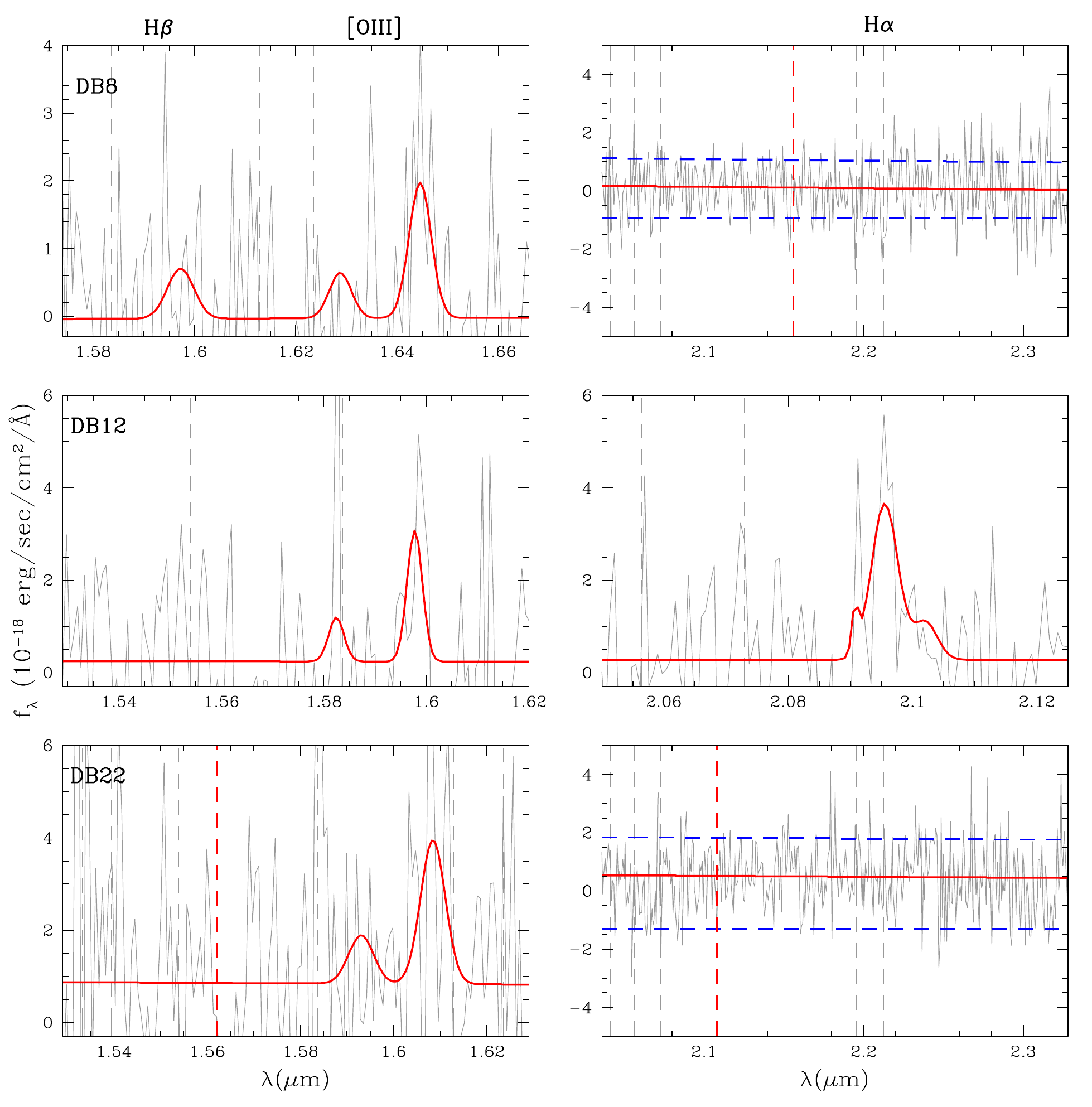}
\includegraphics[width=90mm]{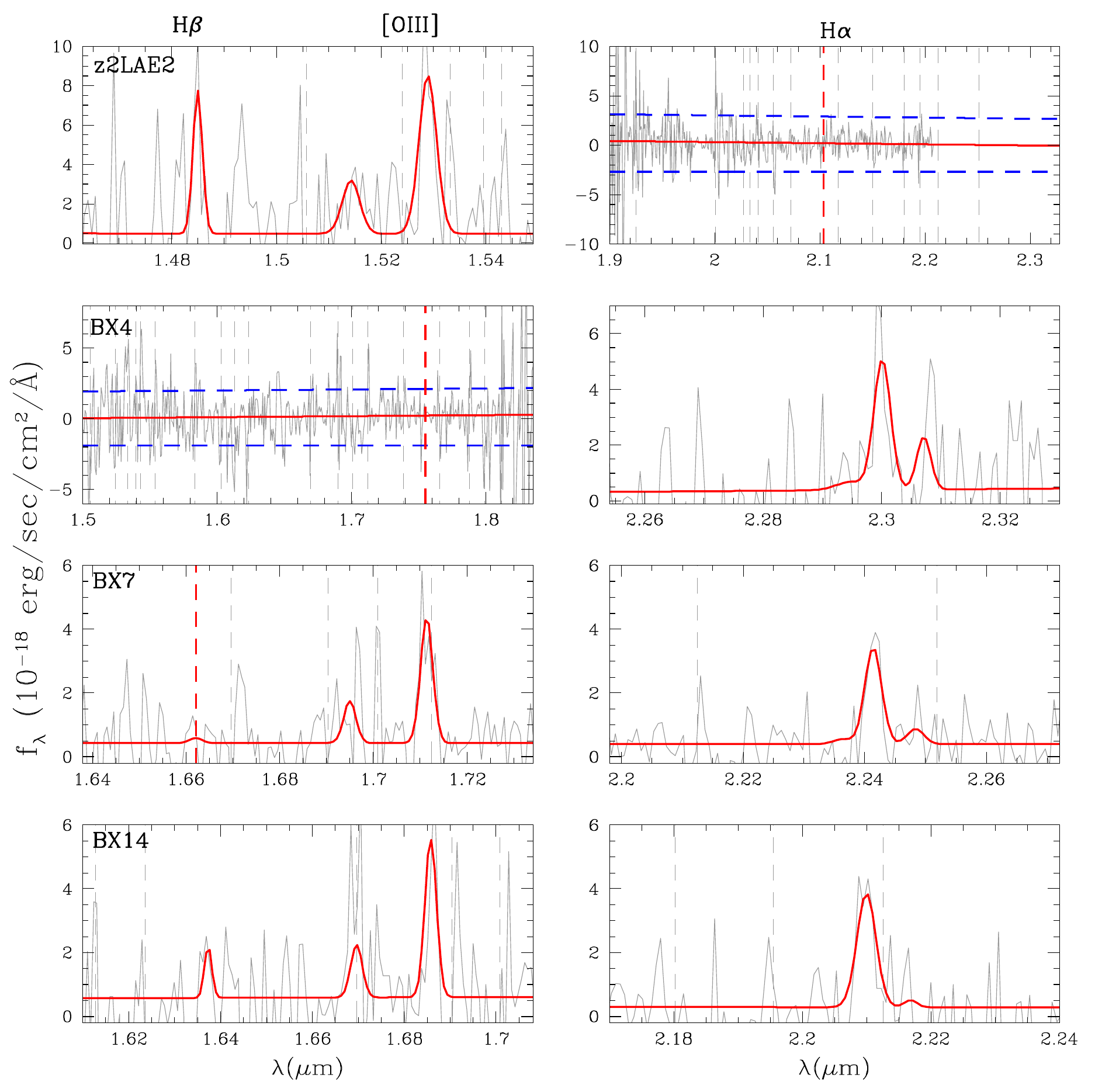}
\caption{Observed 1D spectra (grey curves) of the SFGs detected in our survey, without spectroscopic rest-frame UV counterparts (DB8, DB12, DB22, z2LAE2, BX4, BX7, BX14). [OIII]doublet and H$\alpha$ regions are shown. If only one of these is detected, red vertical dashed lines indicate the location of H$\alpha$ ([OIII]5007) lines which would be implied by the detected-line redshift for DB8, DB22, z2LAE2 (BX4). Dashed grey vertical lines indicate the wavelength of sky emission lines, which could leave residuals in the science spectra. Horizontal dashed blue lines represent the continuum rms in the wavelength range where we did not see an expected emission line. } 
\label{ELother}
\end{figure*}

Among GOODS VIMOS public spectra \citep{Popesso2009, Balestra2010}, we found that BX1 and BX10 were characterized by EW$_{rest-frame}$(Ly$\alpha)>20$ {\AA} and BX16 was characterized by Ly$\alpha$ in absorption. Thanks to MUSYC VIMOS and IMACS rest-frame UV spectra, LAE27 and z3LAE2 were confirmed to be LAEs, while LBG11 showed EW$_{rest-frame} = 8 $ {\AA} (Fig. \ref{ELLAE}). In Table \ref{everypropV} we reported H$\alpha$/H$\beta$, Ly$\alpha$/H$\alpha$, and Ly$\alpha$/[OIII]5007 line ratios for all the SFGs of our survey. The integrated line fluxes were obtained by fitting MMIRS spectra shown in Fig. \ref{ELLAE} and \ref{ELother}. 

For the objects with available rest-frame UV spectra, a good (stellar) reddening estimation could be obtained from the $\beta$ slope of he rest-frame UV continuum. 
This is shown as E(B-V)$_{\beta}$ in Table \ref{everypropV}. Following \citet{Calzetti2000}, $\beta$ is related to E(B-V) through the following equations,

\begin{equation}
A(1600) = 4.85 + 2.31 \times ~ \beta, \\
\end{equation}

\begin{equation}
A(1600) = 4.39 \times Eg(B-V), \\
E(B-V)=c \times Eg(B-V), \\ 
\end{equation}

where A(1600) corresponds to the galactic absorption in magnitude at $\lambda_{rest-frame}=1600$ {\AA}, Eg(B-V) is the reddening of nebular gas, and E(B-V) that of stellar continuum. The constant $c$ corresponds to a factor $0.44 \pm 0.03$ in \citet{Calzetti2000}. However, the Calzetti law was calibrated at low redshift and there are recent works showing that factor could be just $c=1$ at $z\sim2$ \citep{Erb:2006}. \citet{Hayes2010} assumed $c=1$ as well, because they noticed that $c=0.44$ significantly over-predicts the SFR(H$\alpha$) with respect to SFR(UV) for their sample. For our sample, SFR(H$\alpha$) and SFR(UV) are within a factor of 2. From the SED fitting results in Gu2011, we also noted that $0.44<c<1$ seemed to be more appropriate to reproduce SFRs, but we could not estimate one precise value. It is beyond the scope of this paper to investigate if $c$ is closer to 0.44 or 1, so we assumed the Calzetti law entirely. Of course to dust-correct emission line fluxes we used Eg(B-V). 

An alternative way to estimate dust reddening is the Balmer decrement \citep[F(H$\alpha$)/F(H$\beta$)][]{MM1972}. However, the low signal-to-noise H$\beta$ line makes this an estimation which is less reliable than the previous ones. 
We reported the Balmer decrements and the implied stellar color excess, E(B-V)$_{\beta-\alpha}$, for 4 sources in Table \ref{everypropV}.

We considered the stellar dust reddening from stacked SED fitting as the most robust measure to use for dust reddening correction, when a rest-frame UV fit could not be performed. Therefore, we referred to the sub-sample definitions in \citet{Lai:2008} and Gu2011 to infer stellar mass and E(B-V) from their SED fits. The categories, presented in those works and considered for the current analysis are IRAC-faint (f$_{3.6 \mu m}<0.3 \mu$J), IRAC-bright (f$_{3.6 \mu m}\geq0.3 \mu$J), UV-faint ($R\geq25.5$), UV-bright ($R<25.5$) LAEs, `BX' and LBGs. 
The $z\sim2$ LAEs detected in our MMIRS survey all had SEDs consistent with those of UV-bright LAEs (log(M/M$_{\odot}$)=9.1[8.9-9.4], where square brackets correspond to the 68\% confidence level range of stellar-mass values). The ones at $z\sim3$ all belonged to the IRAC-bright sub-sample (M$_* \sim 10^9-10^{10}$ M$_{\odot}$).
The stacked sub-sample SED stellar mass was then rescaled to match the $R$ band magnitude of individual sources. The SED for individual objects, performed in \citet{Hayes2010}, addressed stellar mass and reddening of DB sources. 
In Table \ref{everypropV} we reported the physical parameters from the SED fitting as E(B-V)$_{SED}$ and log(M$_{*}$/M$_{\odot}$). Their error bars accounts for the spread in the sample population (Gu2011).

To infer star formation rate (SFR) of SFGs and LAEs, we applied the \citet{Kennicutt:1998} equation. 
\begin{equation}
      F_{corr}(H\alpha) =  F_{obs}(H\alpha) \times 10^{0.4 \times Eg(B-V) \times K_{Calzetti}(\lambda_{H\alpha})} \\
\end{equation}
\begin{equation}
      SFR_{corr}(H\alpha) =  7.9E-42 \times F_{corr}(H\alpha) \times 4\pi D_{L}(z)^2 \\
\end{equation}
Only for three sources (BX1, BX10, BX16) we were able to estimate E(B-V)$_{\beta}$ (and so Eg(B-V)$_{\beta}$), we used to estimate SFR$_{corr}$(H$\alpha$). On the other hand H$\alpha$ flux was an upper limit for the BX10 spectrum and it translated into an upper limit of SFR$_{corr}$(H$\alpha$). In the case of BX4, BX7, BX14, and DB12, the SFR(H$\alpha$) was assumed to be a lower limit of the SFR$_{corr}$(H$\alpha$) given the uncertain correction factor. On average SFR$_{corr}$(H$\alpha$) was found to be within a factor of two of the value estimated from the rest-frame UV. 

As an example of the quantities introduced above, we present here the calculation relative to BX1.
It was characterized by a quality A spectrum in GOODS survey. Fitting the spectroscopic rest-frame UV continuum with a power law, f$_{\lambda} \sim \lambda^{\beta}$, we obtained $\beta_{best~ fit}=-1.6\pm0.11$ and so E(B-V)$_{\beta}=0.12\pm0.02$, Eg(B-V)$_{\beta}=0.27\pm0.04$.
The Balmer decrement is measured as F(H$\alpha$)/F(H$\beta$) = 2.6$\pm$2.5 i.e. F(H$\alpha$)/F(H$\beta) \leq 5.1$, which implies E$_{\beta-\alpha} \leq$ 0.63. Its stellar reddening is then E(B-V)$_{\beta-\alpha} \leq 0.54$, roughly consistent with the SED estimation, but not very conclusive.

BX1 was originally selected to have $R=24.65$ and it belonged to the UV-bright LAE sub-sample (the $R$ band magnitude of the stacked UV-bright sub-sample SED is 25.0). We, therefore, assumed the stacked SED shape was representative of this object, but we rescaled by a factor 1.4 the stellar mass to account for the individual object having a brighter $R$ mag, yielding log(M/M$_{\odot}$)=9.25[9.05-9.55].

Assuming E(B-V)$_{\beta} = 0.12\pm0.02$ and a Calzetti law for dust attenuation, we calculated F$_{corr}$(H$\alpha$) = (9.4$\pm$2.1)E-17 erg sec$^{-1}$ cm$^2$ and SFR$_{corr}$(H$\alpha$) =  28.5$\pm$8.2 M$_{\odot}$ yr$^{-1}$. This value was roughly consistent with the SFR averaged over the last 100 Myr found in Gu2011 for UV-bright LAEs ($<SFR>_{100}$=10[6-70] M$_{\odot}$ yr$^{-1}$). Following the \citet{Kennicutt:1998} equation for the SFR implied by the ultraviolet continuum, BX1 is characterized by an SFR$_{corr}$(UV, $R$ band) around 50 M$_{\odot}$/yr.

\subsubsection{Line velocity offset}
\label{offset}

In the attempt to measure SFG properties we also studied ISM kinematics. 
[OIII]5007 can be used as a proxy of the galaxy systemic redshift \citep{McLin2011, Fin2011}. Fitting Ly$\alpha$ (from either GOODS or MUSYC spectra) and [OIII]5007 (from MMIRS spectra) emission lines with Gaussian curves we obtained best fits for their central wavelengths.
To compare GOODS with our redshifts, we first applied the heliocentric correction at the time of the observations. It was v=+22.2 km sec$^{-1}$ at Cerro Paranal and v=+16.7 km sec$^{-1}$ at Las Campanas for BX1, as an example. Then, the offset velocity is estimated with the following equation (Table \ref{everypropV}),
\begin{equation}
      \Delta v_{Ly\alpha-[OIII]5007} = c \times \left(\frac{1+z_{Ly\alpha}}{1+z_{[OIII]}}-1 \right) \\
\end{equation}
where rest-frame wavelengths are $\lambda_{Ly\alpha}=1215.67$ {\AA} and $\lambda_{[OIII]5007}=5008.239$ {\AA}.

\subsubsection{Gas phase metallicity}
\label{Met}

In Table \ref{tab:MetEst} we present the line ratios used to infer metallicity, parametrized by 12+log(O/H). We calculated 
[OIII]5007/H$\beta$, [OIII]5007/[NII]6584, and N2 ([NII]6584/H$\alpha$), which relate to metallicity through the \citet[][their Fig. 8 and 17]{Nagao2006} and \citet{Maiolino2008} calibrations. [OIII]5007/[NII]6584 is sensitive to reddening, so it has been corrected using either E$_{\beta}$(B-V) or E$_{SED}$(B-V) values  (when the rest-frame UV spectrum is not available) and the Calzetti law (Table \ref{everypropV}).
We reported [NII]6584/H$\alpha$ for six sources, but only for three of them (BX16, BX4, DB12) the ratio is bigger than its error. It is expected for SFGs with negligible or no AGN contribution that [NII] is a weak emission line, hence it is difficult to use it to constrain metallicity. Therefore, in addition to N2, we also determined metallicity from two other emission line calibrations.    
The two rows in the third column of Table \ref{everypropV} represent the values inferred by the two branches of [OIII]5007/H$\beta$ vs 12+log(O/H) relation. For comparison with the literature we also adopted the \citet{PP2004} calibration of N2 (second row of 7th column) and O3N2 ([OIII]/H$\beta$ / [NII]/H$\alpha$). In the last column we show as 12+log(O/H)$_{adopted}$ the value we will use in the discussion section. 
Due to low signal-to-noise from most [NII] and H$\beta$ emission lines, we were only able to constrain 12+log(O/H) only for two sources (BX16 and DB12) and to set lower (upper) limits for two (two) others. In the case of LBG11, [OIII]5007/H$\beta = 2.8\pm2.4$ gave 12+log(O/H)$\leq9.0$, which is not informative. In the case of z3LAE2, [OIII]5007/H$\beta = 1.7\pm0.8$ was consistent with 12+log(O/H) of [7.0-7.2] and [8.5-8.7]. We considered the upper branch of the [OIII]5007/H$\beta$ calibration the most representative of this source metallicity, because EW(H$\beta$) was measured to be at least 10 {\AA} \citep{HuCowie2009}.

In the range of line ratios we measured for our sample, N2 calibration from \citet{PP2004} and \citet{Nagao2006} agreed within the error bars. As an example, [NII]/H$\alpha=0.2$ for DB12. This ratio corresponds to 12+log(O/H)=8.7 or 8.5 following \citet{Nagao2006} or \citet{PP2004} respectively. This difference of 0.2 dex was within the line ratio error bar. 

As shown by \citet{kewley2008}, metallicity calibrations based on different approaches (such as line ratios and direct measurement of electron temperature) could produce a systematic deviation of 12+log(O/H) up to 0.7 dex. However, the analysis of \citep[][their Fig. 2]{kewley2008} showed that N2 and O3N2 have the lowest deviation from all the other possible calibrations at $9<$ log(M$_*$/M$_{\odot}) <10$, which is the range of masses covered by the sources in our sample. 

We chose recent literature estimations of  metallicity of SFGs and LAEs to compare with in a meaningful way. 

\citet{Atek2011} used a direct method, measuring [OIII]4363 (dependent on electron temperature) versus [OIII]5007 to infer metallicity of line emitter sources at $0.35<z<2.3$. This method is successfully applied to derive low-metallicities ($7.1<$ 12+log(O/H) $<8.3$), because at higher metallicities [OIII]4363 becomes too weak to be measured. They discovered a very metal-poor source at $z=0.7$ with 12+log(O/H) $=7.47\pm0.11$ and log(M$_*$/M$_{\odot}$)$<8$. Here, we took this source as an example of a very metal-poor line emitter. We also compared with \citet{Xia2012}, which used the R23 (([OII]3727+[OIII]5007)/H$\beta$) method to estimate metallicity of SFGs with strong emission lines at $z>1$. 

In addition to them, we considered sources at $z\geq2$. The star-forming galaxy, named BX418, was found to have $12+$log(O/H) $=7.9\pm0.2$ \citep[][direct method]{Erb2010}. With EW(Ly$\alpha$) = $54.0\pm1.2$ {\AA}, it is one of the most metal-poor LAEs known at high redshift. \citet{Fin2011} placed upper limits on metallicity for two LAEs at $z=2.3$ and 2.5, using N2 and O3N2 from \citet{PP2004}. They can be directly compared to our sources. \citet{Nakajima2012a} selected a sample of LAEs at $z\sim2.2$ and were able to set a lower limit of 12 + log(O/H) $\geq$ 7.9 on the stacked object, using a combination of [OII]/H$\beta$ and N2 calibrations. \citet{Nakajima2012b} used N2 from \citet{Maiolino2008} to infer 12+log(O/H) of individual sources, therefore their results can also be directly compared with ours. 

We could apply neither R23 nor direct methods because in our MMIRS spectra we did not cover [OIII]4363 or [OII]3727. 

\begin{figure}[h!]
\centering
\includegraphics[width=70mm]{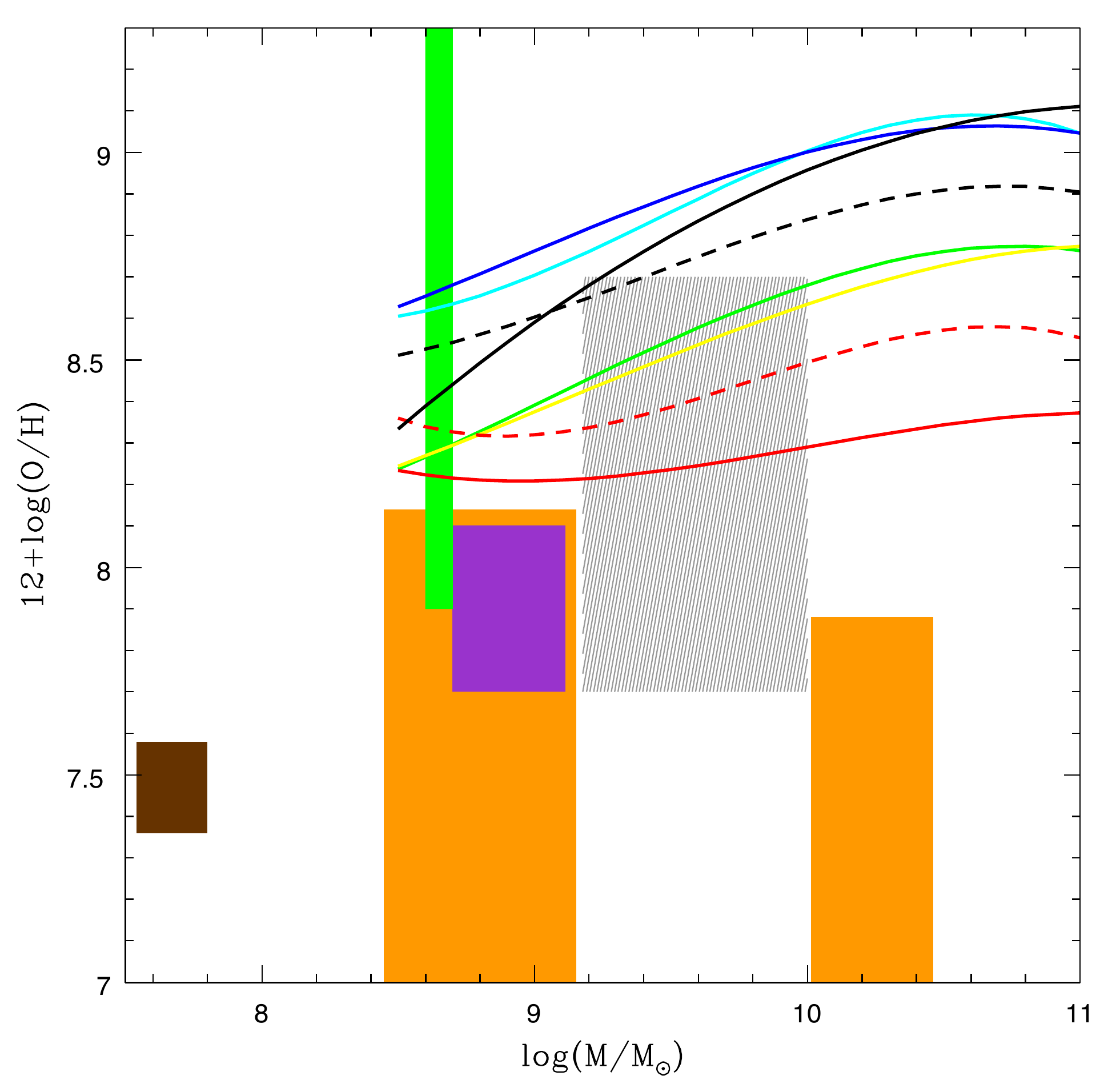} 
\caption{Mass-metallicty relations as presented in \citet{kewley2008} for log(M/M$_{\odot})>8.5$. The color-coding is the same as in their Fig. 2. The cyan curve is obtained from the R23 calibration, blue, dashed black, and dashed red from a combination of R23 and [OIII]/[OII], solid black from a combination of line ratios, such as [OII], H$\beta$, [OIII], H$\alpha$, [NII], [SII], solid green from O3N2, solid yellow from N2, and solid red from the measurement of the electron temperature, using [OIII]4363, 4990, 5007. Shaded regions indicate the mass-metallicity area occupied by literature results within the error bars. Also shown is the \citet{Atek2011} source in brown at the lowest mass range, \citet{Fin2011} upper limits in orange, \citet{Nakajima2012a} lower limit in green, and the range of objects from \citet{Xia2012} $z>1$ in grey. The sources in our survey are characterized by $9<$ log(M/M$_{\odot}) <10$, where the N2 and O3N2 agree with all the other calibrations within 0.4 dex.}   
\label{MetEstfig}
\end{figure}

In Fig. \ref{MetEstfig} we show the mass-metallicity (M$_*$-Z) relations presented in \citep[][their Fig. 2]{kewley2008}, with their same color coding. The shaded regions represent the ranges in stellar mass and metallicity for the literature results described above. 
\citet{kewley2008} pointed out that their M$_*$-Z relation, corresponding to the direct method, was characterized by high statistical  uncertainty. 

At the stellar mass of BX418 (log(M/M$_{\odot}$) $=9.0$), the discrepancy between the direct and N2 methods is less than 0.2 dex. In the range of masses covered by the \citet{Xia2012} sample, it is 0.3 dex between N2 and R23 estimators. 
In the range of masses probed by \citet{Nakajima2012a}, it is less than 0.4 dex among all the calibrations. 
These discrepancies were within the error bars of the metallicity estimations of our sources. Therefore, our results could be compared with the literature ones just mentioned, reasonably without scaling to a common calibration (Sect. \ref{sec:discussion}). 

%%%%%%%%%%%%%%%%%%%%%%%%%%%%%%%%%%%%%%%%%%%%%%%%%%%%%%%%%%%%%%%%%%%%%%%%%%%%
\section{Discussion}  
%%%%%%%%%%%%%%%%%%%%%%%%%%%%%%%%%%%%%%%%%%%%%%%%%%%%%%%%%%%%%%%%%%%%%%%%%%%%
\label{sec:discussion}

We found the [OIII]4959, 5007 emission line doublet to be the strongest feature in MMIRS spectra of $z\sim2-3$ sources. We detected the [OIII] doublet in 12 out of 13 galaxies, while we detected H$\alpha$ in just 6 out of 10 $z\sim2$ galaxies.
This may be surprizing at first glance. However, for SFGs characterized by slightly sub-solar metallicity, 
[OIII]5007 is predicted and measured to be very strong \citep{SB2011}. \citet{kakazu2007} conducted a survey to search for ultra-strong emission line galaxies at $0<z<1$. Their results showed that these galaxies contribute roughly 10\% of the measured SFR density at that epoch.  All the H$\alpha$ emitters in their sample were metal-poor (Z $<0.45$ Z$_{\odot}$) galaxies, characterized by strong [OIII]4363, while [OII]3727 was found to be very weak compared to [OIII]5007.  The higher-metallicity objects had, instead, stronger [OII]3727/[OIII]5007. They concluded that they were looking at galaxies different from local metal-poor dwarf galaxies with larger [OIII]5007/[OII]3727 flux ratio and more mass. This suggested an early chemical enrichment phase in galaxies which are still growing. 
  
Also, in our spectra the sky background continuum can be 3 times higher at 22000 {\AA} than at 16000 {\AA}, and we measured an average fractional background error at 22000 {\AA} twice the one at 16000 {\AA}. 
In Fig. \ref{ELLAE} and \ref{ELother} we presented the spectra of the SFGs in our sample, separated into [OIII]doublet and H$\alpha$ wavelength ranges.

\subsection{Star-forming galaxy evolutionary stage}
\label{stage}

Thanks to our spectroscopic results, we were able to place SFGs in the evolutionary sequence proposed by \citet{Noeske2007b} (SFR$_{corr}$ (H$\alpha$) versus stellar mass relation defined for log(M/M$_{\odot}$) $>9.5$
at $z<2$). It was discovered that SFGs form a distinct sequence with a limited range of stellar masses. This `main sequence' moves as a whole to higher star formation rate as redshift increases. Therefore, galaxies of a given mass tended to present higher SFR at higher redshift; they were much more active on average in the past. One reason is the larger abundance of gas, depleted with time.
For galaxies in the main sequence, the gradual accretion (as opposed to episodic and bursty events such as mergers) is the dominant process, and major mergers are not the dominant trigger of their star-formation activity and evolution. 
The outliers of the mass-SFR relation are galaxies with larger star-formation rate with respect to ordinary SFGs. 

To completely fill this relation, we should be able to estimate SFR(H$\alpha$) for low mass (M$_{*}\leq10^9$ M$_{\odot}$) objects at $z>2$. But the objects we detected in our spectroscopic survey tend to be bright in $K$ band and therefore to be more massive than few $\times$ 10$^9$ M$_{\odot}.$
Ga2007 had first showed that $z\simeq3.1$ LAEs are, typically, 10$^9$ M$_{\odot}$ galaxies in their active phases of star formation and one of the highest specific SFR objects at that redshift. This is in favor of the idea that LAEs were still building stellar mass. If dust is an indication of a more evolved galaxy population and it is one of the reasons why Ly$\alpha$ photons cannot escape a galaxy ISM, we expect LAE to be characterized by lower mass and higher specific star formation rate. \citet{Nilsson:2009} found that $z\sim2$ LAEs instead presented a more massive (log(M/M$_{\odot}$) $\sim9.5$) sub-sample. 

\begin{figure*}
\centering
\includegraphics[width=120mm]{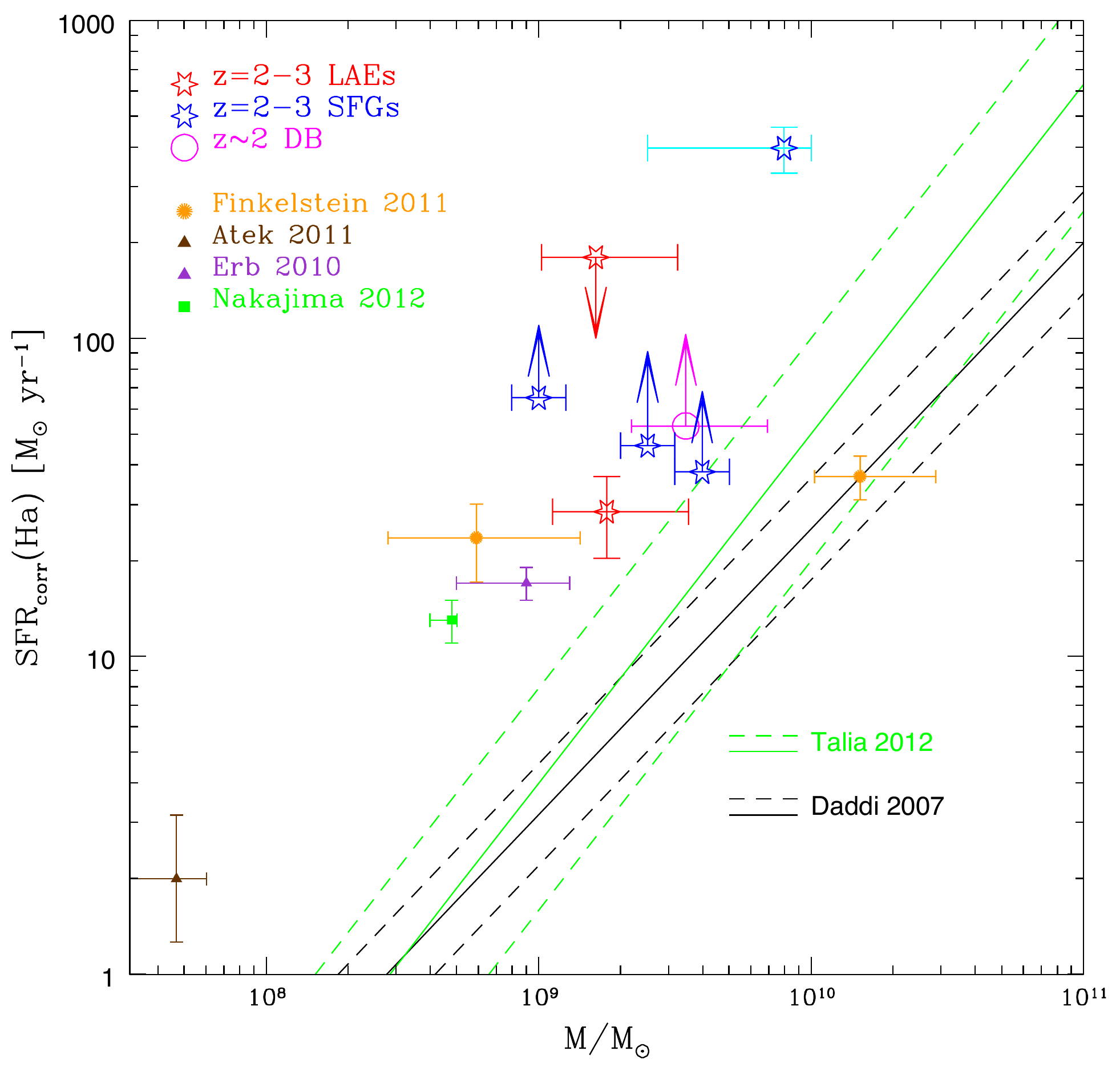}
\caption{SFR$_{corr}$(H$\alpha$) vs stellar mass for LAEs and SFGs at $z\geq2$.
Red and magenta colors indicate Ly$\alpha$ and H$\alpha$ emitting galaxies, while blue indicate UV-continuum-selected SFGs without Ly$\alpha$ in emission. 
The blue star with cyan error bars represents BX16, which shows Ly$\alpha$ absorption. The SFR(H$\alpha$) upper limit is due to an upper limit in H$\alpha$ flux, and lower limits are presented when it is not possible to estimate E$_{\beta}$(B-V) from the rest-frame UV continuum slope for individual sources. The solid green line is the M$_{*}$-SFR relation by \citet{Talia2012} for SFGs at $z\simeq$2. The black line is the same relation for $z\sim2$ SFGs in the GOODS survey (\citet{Daddi2007}). Dashed green and black lines indicate the 1$\sigma$ range of those relations. Measured values from \citet{Nakajima2012a} (green square), \citet{Fin2011} (orange dots), \citet{Atek2011} (brown triangle), and \citet{Erb2010} (violet triangle) are also shown.}
\label{SFRM}
\end{figure*}

In Fig. \ref{SFRM} SFR$_{corr}$ (H$\alpha$) is plotted versus stellar mass for SFGs at $z\sim2-3$. 
Small symbols correspond to the literature data we chose for comparison, as described in the previous section.
The lines (solid) with associated scatter (dashed) represent the correlation derived for $z\sim2$ SFGs in the GOODS field by \citet{Daddi2007} and from GMASS (Galaxy Mass Assembly ultra-deep Spectroscopic Survey) by \citet{Talia2012}.
Our LAE and SFG data points tend to be characterized by higher SFR than typical $z\sim2$ SFGs. \citet{Atek2011} observed that metal-poor galaxies at $z\sim1-2$ are located towards the upper left corner 
of Daddi's correlation, indicating that they are experiencing strong metal-poor burst episodes of star formation. 
The metal-poor source from \citet{Erb2010}, the sources from \citet{Nakajima2012a}, and the lowest mass LAE from \citet{Fin2011} are also located in the same region.

\begin{figure*}
\centering
\includegraphics[width=120mm]{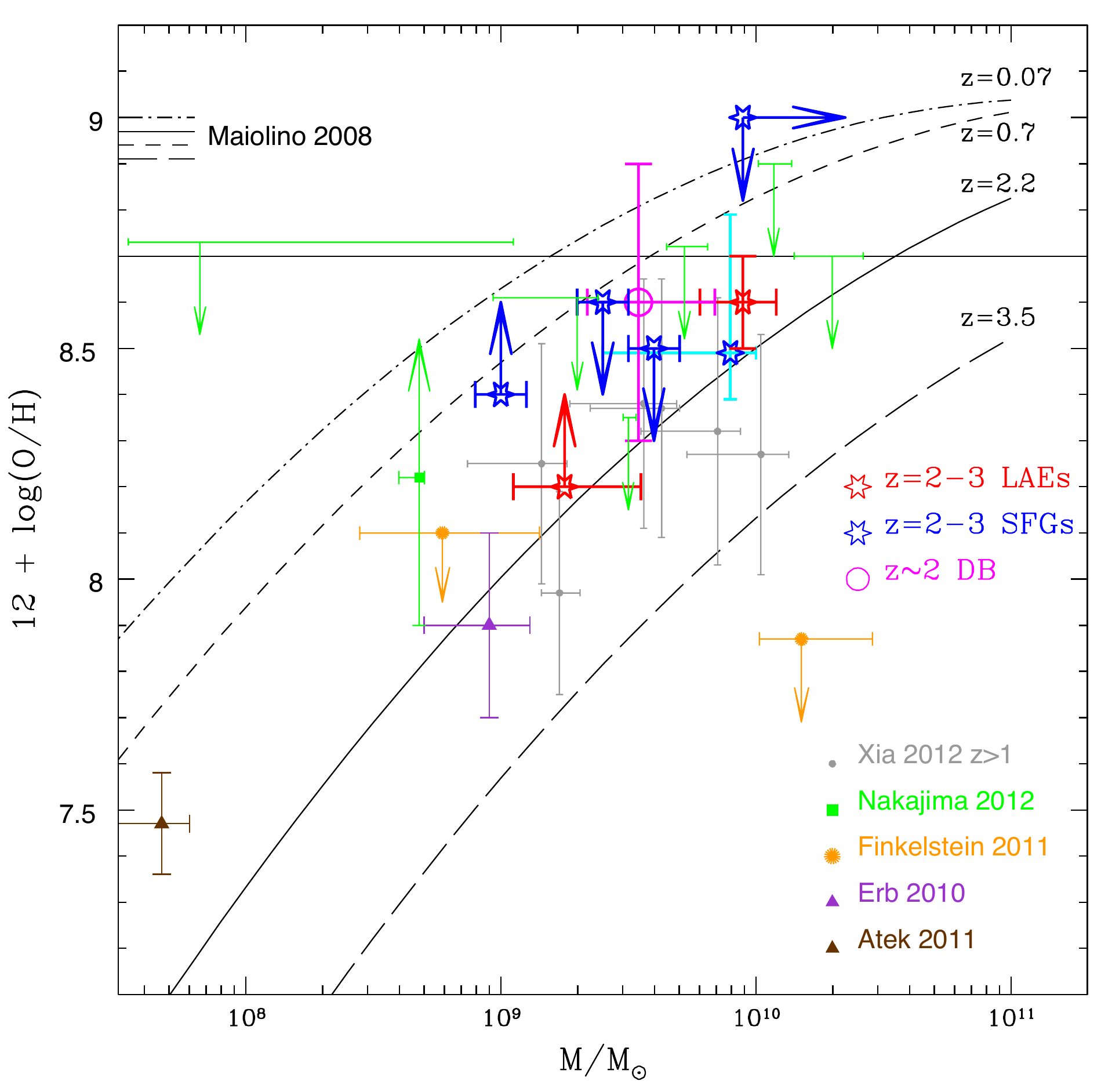} 
\caption{Metallicity estimation versus stellar mass. The big magenta circle, red and blue stars correspond respectively to DB, LAEs and SFRs with EW$_{rest-frame}$(Ly$\alpha)< 20$ {\AA}  in our sample. Thin green upper limits and the green square are the data points from \citet{Nakajima2012b}(2012b), small grey dots are \citet{Xia2012} values for $z>1$ emitting galaxies. Black long-dashed, solid, dashed and dotted-dashed curves are the M$_{*}$-Z relations from \citet{Maiolino2008} for the labeled redshifts. The horizontal solid line indicates solar metallicity (12+log(O/H)=8.7). }
\label{ZM}
\end{figure*}

 In Fig. \ref{ZM} we show metallicity versus stellar mass for our and the literature sources we chose for comparison. 

The sources detected so far in our survey are confirmed to be active star-forming, characterized by metallicity of the order of $0.3<Z/Z_{\odot}<1.2$ (12 + log(O/H) $\sim$ 8.2-8.8). As explained in Sect. \ref{Met}, it is meaningful to compare our results with the ones presented in the figure. The black curves are the mass-metallicity relations from \citet{Maiolino2008}, which used spectroscopic data from AMAZE (Assessing the Mass-Abundance redshift[-Z] Evolution). 
They estimated stellar masses assuming a Salpeter IMF, like in our SED fitting, and calculated the best fit parameters for the relation at z=0.07 (dotted-dashed), z=0.7 (dashed), z=2.2 (solid, calibrated following \citet{Erb:2006} observations), and z=3.5 (long dashed). They showed evolution in the mass-metallicity relation stronger at lower stellar masses. Our six star-forming galaxy metallicity estimations are roughly consistent with \citet{Maiolino2008} M$_*$-Z relation at $z=1-2$. The \citet{Fin2011} source with stellar mass bigger than 10$^{10}$ M$_{\odot}$ results in a very metal-poor object. 

Our UV-bright LAEs do not seem to be the most metal-poor objects at this redshift, even if for a couple of them we just set upper limits. This could suggest an early chemical enrichment during the growth phase of individual $z\sim2-3$ Ly$\alpha$ emitting galaxies \citep{kakazu2007}. The Erb et al. and Atek et al. very metal-poor objects present values even lower than the lower limit inferred by \citet{Nakajima2012a} for $z\sim2.2$ emitters.

The existence of a correlation between stellar mass and metallicity reflects the fundamental role that galaxy mass plays in galactic chemical evolution. It can mean that low-mass galaxies could be already experiencing a chemical enrichment phase. Also, galactic winds could be responsible in removing metals from galaxies \citep{Tremonti2004,Rodrigues2012}. This hypothesis could explain the metallicity lower than 7.9 obtained by \citet{Fin2011} for a rare metal-poor source at that stellar mass. \citet{Tremonti2004} also noted that in large galaxies with deep potential wells, star formation activity and supernova explosions are not effective in pushing material in the form of an outflow. Inflows of metal-poor gas could either dilute the gas, reducing its metallicity, or turn on new star-formation episodes which would again metal-enrich the ISM.
In addition to this, \citet{Mannucci2010} showed that the mass-metallicity relation, observed in the local and high-z Universe, may be due to a more general relation between stellar mass, metallicity and SFR. Based on their results,  $z\sim2.5$ SFGs with log(M$_*$/M${_\odot})=9.7$ are characterized by 12+log(O/H) of the order of 8.5-8.6, for star formation rates of about 30 M$_{\odot}$ yr$^{-1}$. For the same stellar mass and metallicity, the LAEs considered here could present even higher SFR. This can be an indication that different mechanisms could dominate in this kind of galaxies. We investigate the outflow hypothesis in the next sub-section and how it is related to the Ly$\alpha$ luminosity.

We are not yet able to derive conclusions about either M$_*$-Z or M$_*$-SFR relations of LAEs and SFGs at the same redshift. On-coming spectroscopic data are needed to support the hypothesis that $z\sim2$ LAE could be younger but more metal-rich than $z\sim3$ LAEs, as found by \citet{Acquaviva2012}.

\subsection{Ly$\alpha$ emitter ISM properties}
\label{sec:LAE}

\begin{figure*} 
\centering
\includegraphics[width=120mm]{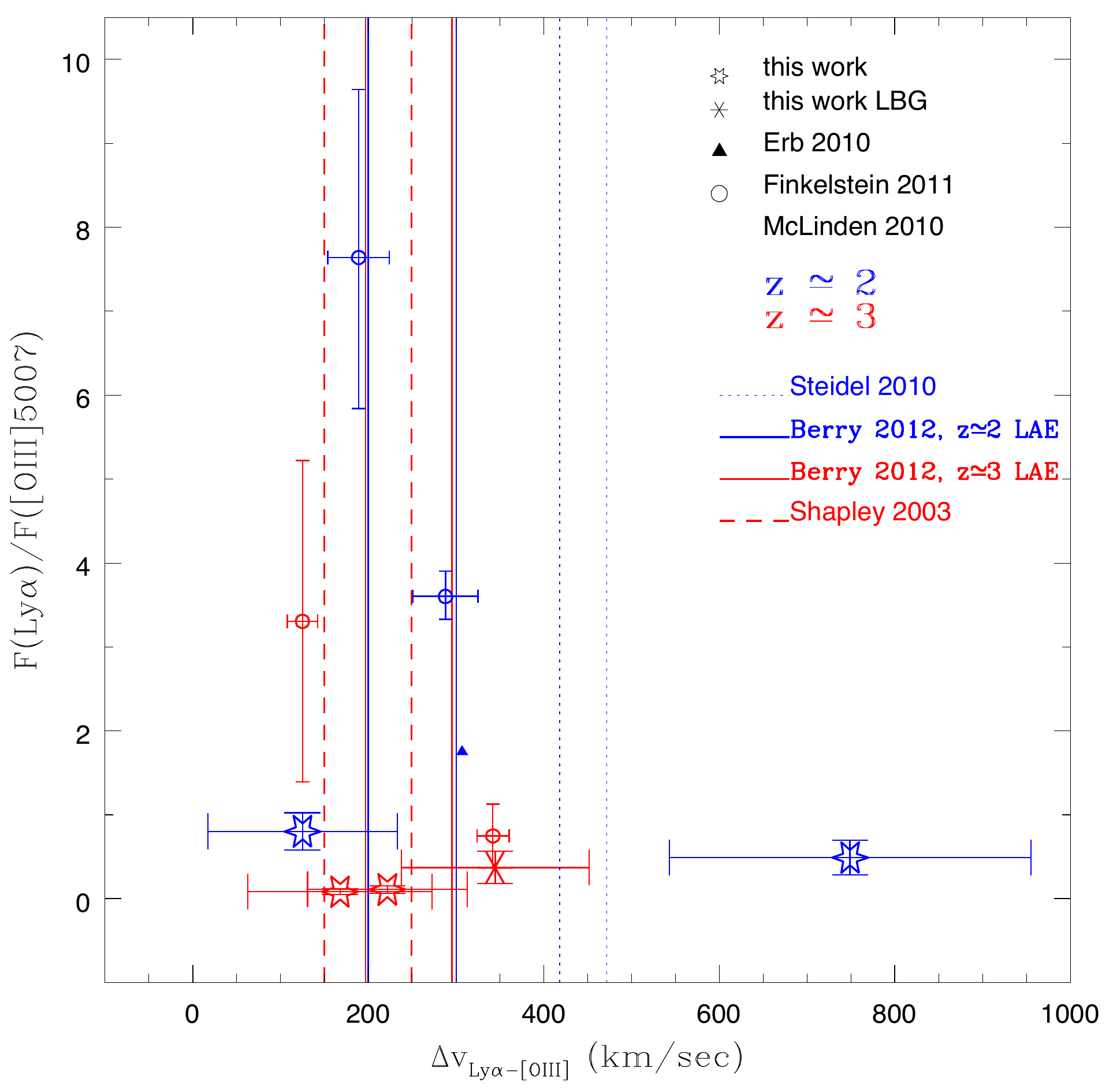} 
\caption{F(Ly$\alpha$)/F([OIII]5007) ratio versus the velocity offset between the same emission line central wavelengths. We present LAEs at $z\simeq2$ in blue and at $z\simeq3$ in red. The red asterisk is for LBG11. Data from \citet{McLin2011} and \citep[][updated by Chonis et al. private communication]{Fin2011} are shown as circles, from \citet{Erb2010}  as triangle. The vertical lines represent the ranges of $\Delta v_{Ly\alpha-LIS}$ by \citet{Shapley:2003} (dashed) at $z\sim3$, \citet{Berry2012} (solid) at $z\sim2$ and 3, roughly divided by 3 to be compared with $\Delta v_{Ly\alpha-[OIII]5007}$ (see text), and \citet{Steidel2010} (dotted). }
\label{Deltav}
\end{figure*}

From the MUSYC \citep[FORS, VIMOS, and IMACS; see][]{Berry2012} and GOODS-VIMOS spectroscopic surveys, we were able to measure Ly$\alpha$ fluxes (Fig. \ref{Deltav}) for 2 LAEs at $z\sim2$, 1 SF galaxy with Ly$\alpha$ in absorption, 2 LAEs at $z\sim3$, and 1 LBG. GOODS 1D spectra were generated through an optimal extraction by following the slit profile measured in each slit. This minimized the slit losses for individual objects. Also, the GOODS team restricted the observations to airmass equal to 1.1, to minimize the loss of light due to atmospheric refractions \citep{Popesso2009, Balestra2010}.

The Ly$\alpha$/H$\alpha$ flux ratio is generally investigated to study the ISM nature \citep{Hayes2010,Fin2011}. In fact the presence of dust (to which the Balmer decrement is also sensitive) can significantly reduce the escape of Ly$\alpha$ versus H$\alpha$ photons. For BX1 (Sect. \ref{offset}) we calculated an escape fraction of Ly$\alpha$ photons of about 30\% and for BX10 a lower limit of 7\% due to F(Ly$\alpha$)/F(H$\alpha$)  $\geq 0.6$. \citet{Fin2011} calculated a ratio close to 7. First of all, they were observing two LAEs brighter in Ly$\alpha$ than our originally continuum-selected galaxies. Also, besides the big uncertainty and the low statistics, we could be observing individual LAEs with a higher dust amount than theirs.
But they calculated E(B-V) = $0.20\pm0.15$ and E(B-V) $< 0.13$ from the Balmer decrement for their two objects. We estimated E(B-V) = $0.12\pm0.02$ (E(B-V) $< 0.54$) for BX1 and E(B-V) = $0.12\pm0.02$ for BX10 from fitting the UV $\beta$ slope (Balmer decrement), which are consistent with theirs. 
Therefore, in addition to dust reddening, other ingredients of LAE interstellar medium need to be considered to completely understand their properties, such as gas phase metallicity and kinematics. 

For most emission line objects, the Ly$\alpha$/[OIII]5007 flux ratio was calculated with high significance (S/N $\geq3$). 
This ratio is affected by gas phase metallicity, kinematics, and Ly$\alpha$ radiative transfer in the LAE ISM, related to the neutral Hydrogen content.
Because metallicity could be a key ingredient which regulates Ly$\alpha$ photon escape, we note that Ly$\alpha$/[OIII]5007, in addition to the Ly$\alpha$/H$\alpha$ or together with SED fitting, can be investigated to improve the understanding of star-forming galaxy ISM (temperature, metal abundances, geometry, kinematics). 
We noted a tentative trend between Ly$\alpha$ and [OIII]5007 emission lines, which could be related to the M$_*$-Z relation.
High statistics are needed to disentangle Ly$\alpha$/[OIII]5007 dependence on metallicity and dust content. 

We noted that SFGs at $z\sim2-3$ show F(Ly$\alpha$)/F([OIII]5007) ratio between 0 and 1, with $z\sim2$ LAEs showing a ratio closer to 1 (Fig. \ref{Deltav}). In addition to the flux uncertainty reported in Table \ref{tab:ELflux}, the uncertainties described in Sect. \ref{error} are included in the [OIII] flux error budget. 
We used the public version of GalMC code (\citet{Acq2011}), which makes use of Anders\&Fritse 2003 line ratios, to predict Ly$\alpha$/[OIII]5007 ratios of typical SFGs at different values of galaxy age and metallicity (Z). Assuming an age of 10$^8$ yr, for $0.02<$ Z/Z$_{\odot}<2.0$, Ly$\alpha$/[OIII]5007 can be between 5 and 8, assuming case B recombination and no dust. For E(B-V) $=0.1$ it becomes $2.5<$ Ly$\alpha$/[OIII]5007 $<4.0$ and for E(B-V) $=0.3$ we calculated Ly$\alpha$/[OIII]5007 $<1.0$, assuming the Calzetti law.
Also, in the absence of dust [OIII]5007/H$\beta$ $\sim 4$ and [OIII]5007/H$\alpha$ $\sim1.4$.
Therefore, a Ly$\alpha$/[OIII]5007 $\leq1.0$ could be explained either with E(B-V) $\sim0.3$, consistent with the SED fitting results, or involving radiative transfer effects, which preferentially absorb Ly$\alpha$ with respect to H$\alpha$ and [OIII] photons. However, the referred Finkelstein's LAEs present stronger Ly$\alpha$ emission with respect to [OIII] than ours and significantly lower metallicity. 

It was proposed \citep[][as examples]{Kunth1998, Heckman2001, Dijkstra2011} that galaxy outflows, driven by star formation activity, are able to red-shift the central wavelength of the Ly$\alpha$ photon $\lambda$-distribution. This way they appear invisible to neutral Hydrogen and can escape. To investigate the dependence of Ly$\alpha$/[OIII] on kinematics we measured Ly$\alpha$-[OIII] velocity offsets (Sect. \ref{offset}).

Using stacked rest-frame UV spectra of a few LAEs in our sample, \citet{Berry2012} measured velocity offsets of about 600 km sec$^{-1}$ between Ly$\alpha$ and low-ionization absorption lines (LIS). As low-ionization absorption lines are thought to be tracing the neutral gas, this offset is thought to be produced because out-flowing neutral gas allows only the red side of Ly$\alpha$ line photons to escape, making it appear redshifted with respect to the systemic velocity.
Here we estimated expansion velocities of possible outflows calculating velocity offsets between the central wavelength of Ly$\alpha$ with respect to that of metal lines at the systemic redshift.

In Fig. \ref{Deltav} we present the F(Ly$\alpha$)/F([OIII]5007) ratio versus $\Delta v_{Ly\alpha-[OIII]5007}$.  In the figure we do not show the value of $-897.8\pm257.0$ km sec$^{-1}$, calculated for the SFG with Ly$\alpha$ in absorption (BX16). BX10 is characterized by $749.9\pm188.3$ km sec$^{-1}$ and the highest uncertainty in [OIII] central wavelength ($18095.5\pm11.6$ {\AA}). In fact this emission line is on top of sky line residuals (Fig. \ref{ELLAE}), which affect the line integrated flux and central wavelength. However, we can not exclude it being a high-velocity-offset source. For the other three LAEs at $z\sim$ 2-3 (star symbols), the offsets are 70-270 km sec$^{-1}$. These values are consistent with previous estimations from \citet{McLin2011} and \citep[][updated by Chonis et al. private communication]{Fin2011}. For LBG11, we estimated $\Delta v_{Ly\alpha-[OIII]5007} = 344.5\pm107.0$ km sec$^{-1}$, which is consistent with the value estimated by \citet{McLin2011} for their strongest [OIII]5007 emission line LAE (F([OIII]5007)=($35.48\pm1.15$)E-17 erg sec$^{-1}$ cm$^{-2}$) at similar redshift. 
The velocities we calculated here are also consistent with expansion velocities, Vexp = 100-300 km sec$^{-1}$, measured by \citet{V:2008} for a sample of 11 LBGs.
Instead, \citet{Steidel2010} estimated stronger velocity offsets of $445\pm27$ km sec$^{-1}$ for a sample of H$\alpha$ emitters at $z=2.3$. Their range of Vexp within 1$\sigma$ is plotted as a rectangular region delimited by dotted lines. 

While the outflow velocity, Vexp, is best estimated from the velocity difference between LIS, which trace the out-flowing material, and systemic redshift emission lines, $\Delta v_{Ly\alpha-[OIII]5007}$ can give upper limits for commonly assumed geometries.
In \citet{Berry2012} and \citet{Shapley:2003}, they estimated offsets between LIS absorption lines and Ly$\alpha$, which could be up to three $\times$ Vexp. It was shown in \citet{V:2008} that for an expanding shell model $\Delta v_{Ly\alpha-LIS}$ can be 3 (2) times of the expansion velocity, depending on the HI column density, NHI $\geq (<) 10^{20}$/cm$^2$. The situation can change for different geometries. 
In the figure the range of Vexp values from \citet{Shapley:2003} is presented as the rectangle delimited by the red dashed lines, rescaled to be compared to our $\Delta v_{Ly\alpha-[OIII]5007}$. The range of Vexp values from \citet{Berry2012} at $z=2$ and $z=3$ is also rescaled and it corresponds to the rectangle delimited by solid lines.

We fitted Ly$\alpha$ line profiles with single symmetrical Gaussian curves. The asymmetry of Ly$\alpha$ profile of LAEs at $z\simeq3.1$ was quantified by \citet{McLin2011} as the $\sigma_{gauss}$ of the red peak divided by the $\sigma_{gauss}$ of the blue peak ($a$). One of their two sources showed significant asymmetry in the Ly$\alpha$ line profile, $a=2.1\pm0.2$ ($a=1.0\pm0.1$ for the other source). They used the Hectospec multi-fiber spectrograph at the 6.5m MMT observatory, with an instrument resolution of $\sim$6 {\AA}. We performed a test to estimate the centroid shift due to the symmetric versus asymmetric profile fit. A good fit to the data can be obtained with a symmetric profile and a central wavelength shift of about 1 {\AA} in the rest frame. It implies a velocity offset of 60-80 km sec$^{-1}$ for LAEs at $z\sim3-2$, which is within the error bars of our measurements.  

We do not see any significant difference between redshift 2 and 3, but the statistics are still too low to derive strong conclusions. However, combining Fig. \ref{ZM} and \ref{Deltav} we can observe that lower dust, lower metallicity, bright Ly$\alpha$ flux sources, such as the two LAEs studied by \citet{Fin2011}, tend to be characterized by velocity outflows similar to those of our sources. The rare metal-poor object by \citet{Fin2011} could host an episode of chemical enrichment of pristine gas, also characterized by a low amount of dust. The sources in our survey could be galaxies in more evolved stages, but with hundred km sec$^{-1}$ outflows as channels for Ly$\alpha$ photon escape.

\begin{figure*}
\centering
\includegraphics[width=120mm]{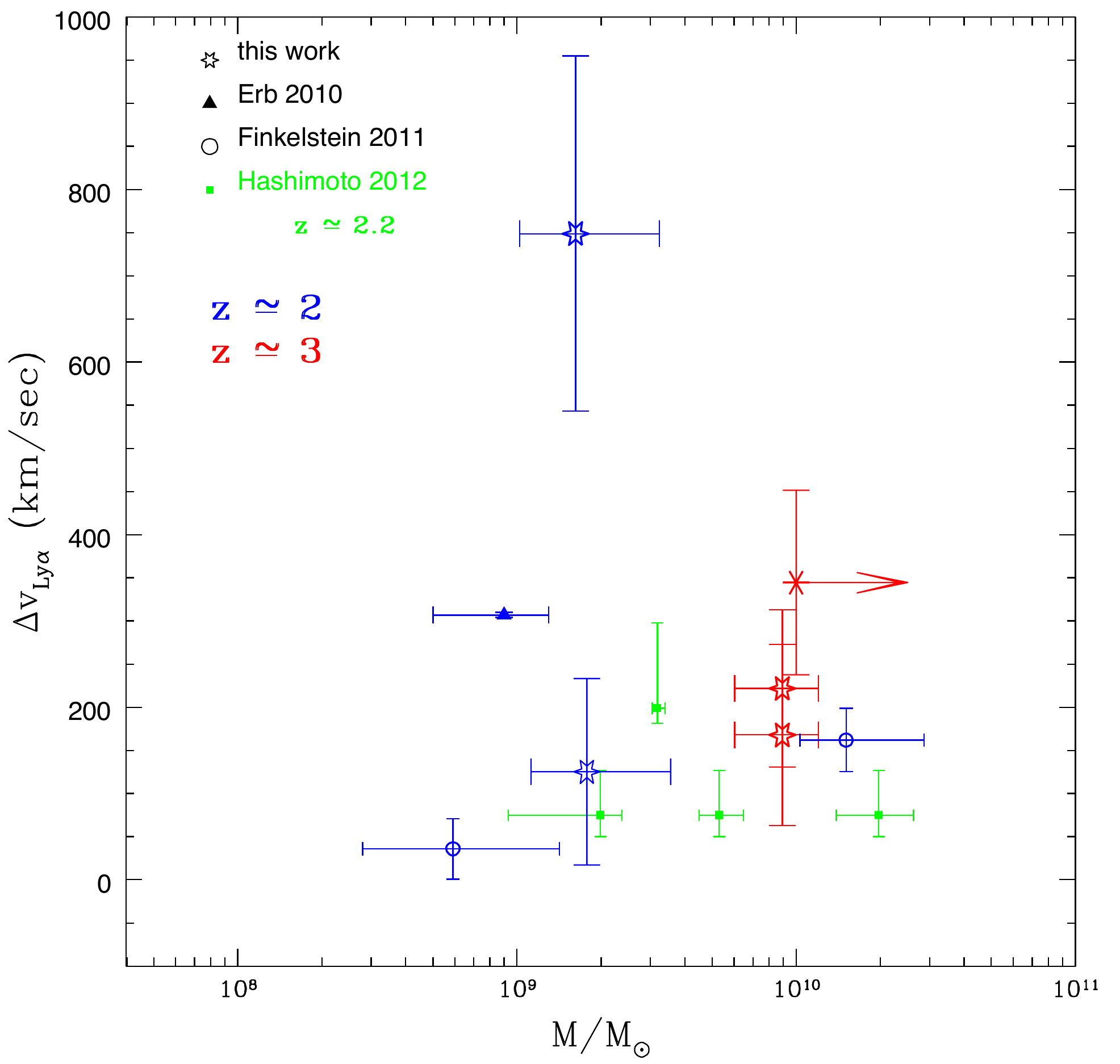}
\caption{Velocity offset between Ly$\alpha$ central wavelength and systemic redshift versus stellar mass. The color coding is as in Fig \ref{Deltav}. Green squares are from \citet{Hashimoto2012} for the sample of $z\sim2.2$ LAEs, also studied in \citet{Nakajima2012b}. The stellar masses of our sources are estimates from stacked SED fitting as explained in Sect. \ref{SEDpro}. }
\label{Mv}
\end{figure*}

We are also interested in understanding how galaxy physical properties can relate to outflow expansion velocities. In Fig. \ref{Mv} we plot $\Delta v_{Ly\alpha-[OIII]5007}$ versus stellar mass for our survey of $z\sim2-3$ SFGs and recent results described in the literature. The squares correspond to \citet{Hashimoto2012}, who fitted individual galaxy SEDs to obtain stellar mass and used the MMIRS and Keck/NIRSPEC instrument to measure $\Delta v_{Ly\alpha-[OIII]5007}$, $\Delta v_{Ly\alpha-H\alpha}$, and to infer Ly$\alpha$ velocity offsets. A weak trend of velocity offset with stellar mass can be seen, but individual cases of very high expansion velocity and log(M$^*$/M${_\odot}$)$\sim9.0$ need to be investigated in more detail.
This trend could be explained considering that low mass galaxies are still in the process of building up their mass and have a weak gravitational potential to support strong outflows \citep{Heckman2001, Orsi2012}. In this scenario log(M$^*$/M${_\odot}$)$=10$ SFGs, like LBGs, could show higher velocity outflow. LBG11, for example, shows one of the highest Vexp of our sample. However, massive galaxies can be characterized by high HI column density (NHI $\sim$ 10$^{20}$/cm$^2$), which would 
completely remove the Ly$\alpha$ line blue peak and would leave a Ly$\alpha$ emission line profile either asymmetric or double-peaked with the red side stronger than the blue one \citep{V:2008, Orsi2012}. This would produce a big $\Delta v_{Ly\alpha-[OIII]5007}$, without the need of strong Vexp. Multiple scattering can produce multiple peaks in the wings of the double-peaked profile \citep{Kulas2012}. 

Additional investigation of Ly$\alpha$ line profile needs to be carried out to disentangle these hypotheses.

%%%%%%%%%%%%%%%%%%%%%%%%%%%%%%%%%%%%%%%%%%%%%%%%%%%%%%%%%%%%%%%%%%%%%%%%%%%%
\section{Summary and conclusions} \label{sec:conclusion}
%%%%%%%%%%%%%%%%%%%%%%%%%%%%%%%%%%%%%%%%%%%%%%%%%%%%%%%%%%%%%%%%%%%%%%%%%%%%

In this paper we presented new spectroscopy of SFGs at $z\sim 2-3$ covered by MUSYC, probing nebular emission. 
The original aim was to detect [OIII] and H$\alpha$ emission lines in SFGs to better understand their ISM properties (Table \ref{tab:MetEst}) and to relate them to Ly$\alpha$ emission line strength and profile, where observed. Details on MMIRS performance are presented in the Appendix.  In Fig. \ref{ELagn}, \ref{ELLAE}, \ref{ELother} we showed the extracted and fitted 1D spectra for all the sources in our survey.
In Table \ref{tab:ELflux}, \ref{tab:MetEst} and \ref{everypropV} we presented measured line fluxes, flux ratios and galaxy physical properties. We list here the main conclusions of our NIR survey.

1) We developed a reduction pipeline, which makes use of COSMOS software. We found COSMOS plus an $ABBA$ source offset procedure successful in reducing MMIRS data (Sect. \ref{sec:red} and details in the Appendix A).
We tested MMIRS behavior using high signal-to-noise spectra of AGNs in our masks. 

2) Through MMIRS spectra, we could confirm seven redshifts obtained via rest-frame UV spectra and determine seven(three) new redshifts of SFGs (AGNs). 

3) We estimated optical emission line intensity and line ratios for 13 SFGs, 6 of which were previously selected as either Ly$\alpha$ and/or H$\alpha$ emitters. 
Two additional SFGs were discovered to be LAEs in GOODS public survey spectra. We found the [OIII]5007 emission line to be the most prominent one in our $z\sim2-3$ galaxy spectra, which are all characterized by R $\leq$ 25.5.  

4) We adopted either rest-frame UV $\beta$ slope or E(B-V)$_{SED}$ for reddening correction. We took advantage of the SED fitting results from Gu2011, \citet{Lai:2008} and \citet{Hayes2010} to estimate the stellar mass of the galaxies in our sample.
In the M$_*$-SFR(H$\alpha$) plane (Fig. \ref{SFRM}), the SFGs in our sample were located towards the upper left corner of the correlation found for $z\sim2$ SFGs by \citet{Daddi2007} and \citet{Talia2012}. At fixed stellar mass, they presented larger SFR than typical SFGs. This implies that they are experiencing active bursts of star formation, which are building up their mass. One possibility can be the star-formation episodes implied by mergers \citep{Noeske2007b}. As observed by \citet{Atek2011}, low-metallicity sources occupy the same locus of the M$_*$-SFR(H$\alpha$) plane. We estimated metallicity from Nagao (2006) and Maiolino (2008) [OIII]5007/H$\beta$, [OIII]5007/[NII]6584, and [NII]6584/H$\alpha$ calibration. 
We chose a few metal-poor sources from the literature to compare our metallicity estimations and demonstrated it is meaningful to compare them. Due to weak H$\beta$ and [NII] emission lines, we could estimate metallicity for two SFGs (one of which is selected to be H$\alpha$ emitter) and set limits for another four.  At $9<$ log(M$_*$/M${_\odot}$) $<10$ our sources have metallicity consistent with 12 + log(O/H) $\sim 8.2-8.8$. The two LAEs studied by \citet{Fin2011} were significantly brighter in Ly$\alpha$ (3-6 times in the integrated line flux, Fig. \ref{Deltav}), showed significantly lower metallicity, and had average lower SFR$_{corr}$(H$\alpha$) than ours. In the M$_*$-Z plane, our sources agreed with the relations found by \citet{Maiolino2008} for $z\leq2.2$ galaxies. 
The sources detected so far in our survey are confirmed to be actively star-forming, characterized by metallicity of the order of $0.3<Z/Z_{\odot}<1.2$. However, we are not yet able to derive strong conclusions about the connection between M$_*$-Z and M$_*$-SFR relations of LAEs and SFGs at the same redshift.

5) We were able to study nebular emission from four LAEs, one SFG with Ly$\alpha$ in absorption, and one LBG with EW$_{obs-frame}$(Ly$\alpha$) = 34 {\AA} (EW$_{rest-frame}$(Ly$\alpha) \sim 2$ {\AA}).
For high-redshift galaxies with both rest-UV and rest-frame optical spectroscopy, our observations indicate that the two emission lines with highest signal-to-noise will often be Ly$\alpha$ and [OIII]5007. For many dim objects, these will be the only lines detected.  Hence, we find empirical motivation to form and study the Ly$\alpha$/[OIII] flux ratio and to attempt to calibrate it both empirically and theoretically.  This ratio is far from ideal as a physical tracer, being sensitive to a combination of metallicity, ionization, and Ly$\alpha$ radiative transfer in the ISM, and it will be challenging to model.  But given the arrival of a generation of multi-object NIR spectrographs and the difficulty of detecting H$\alpha$ for significant redshift ranges of interest, the Ly$\alpha$/[OIII]5007 ratio is worthy of investigation.

Consistent with previous estimations $\Delta v_{Ly\alpha-[OIII]5007}$ was positive, indicating the presence of outflows. 
We found that LAEs are characterized by Vexp $\sim$ 70-270 km sec$^{-1}$, in agreement with \citet{Berry2012}. 
However, the scattering of Ly$\alpha$ photons due to neutral Hydrogen atom, could mask the exact value of the outflow velocity.
It was proposed by \citet{Tremonti2004} that outflows could remove metals together with neutral Hydrogen.  Therefore, high-stellar-mass galaxies could have experienced outflow phenomena in the past and now show lower metallicity. The outflow phenomena could still be going on. However, our sources present just slightly sub-Z$_{\odot}$. 
In Fig. \ref{Mv} we showed velocity offset versus stellar mass. There is a weak trend for which higher mass SFGs could be characterized by higher velocity outflow. However, massive SFGs tend to be characterized by NHI $\sim$ 10$^{20}$/cm$^2$, which could produce a strongly red-shifted red Ly$\alpha$ peak without the need for strong Vexp.

The MMIRS spectrograph can be successfully used to study nebular emission of bright SFGs at $z\sim2-3$, but a larger sample of galaxies is needed to derive strong conclusions.

%%%%%%%%%%%%%%%%%%%%%%%%%%%%%%%%%%%%%%%%%%%%%%%%%%%%%%%%%%%%%%%%%%%%%%%%%%%%
\begin{acknowledgements}
We are grateful for support from the following NSF grants:10-55919, 08-07570, 08-07885; from Basal-CATA PFB-06/2007 (HF, FEB, NP), CONICYT-Chile under grants FONDECYT 1101024 (FEB), ALMA-CONICYT 31100004 (FEB), and ANILLO ACT1101 (FEB), and {\it Chandra} X-ray Center grant SAO SP1-12007B (FEB); from CONICYT-Chile under grant FONDECYT  3100025, Comité Mixto ESO-Chile and FONDAP Center for Astrophysics 15010003 (HF); from Fondecyt Regular 1110327 (NP).
We gratefully thank CNTAC for their generous support of our observing programs CHILE-2010B-0046 and CHILE-2011A-0002.
We acknowledge Brian McLeod and Paul Martini for their strong help given during the reduction process of MMIRS data, Francesco di Mille and Vincent Suc for helping during the observations, and Anthony Gonzalez, Mike Berry, and Robin Ciardullo for useful comments that improved this paper. LG thanks Peter Kurczynski, Peter Laursen, Erik Zackrisson and Florent Duval for useful discussions.
\end{acknowledgements}

%%%%%%%%%%%%%%%%%%%%%%%%%%%%%%%%%%%%%%%%%%%%%%%%%%%%%%%%%%%%%%%%%%%%%%%%%%%%
%% references
%\bibliographystyle{aa}    %% bibliography style file aa.bst from A&A
%\bibliography{biblio}        %% your Bibtex entries copied from ADS

%%%%%%%%%%%%%%%%%%%%%%%%%%%%%%%%%%%%%%%%%%%%%%%%%%%%%%%%%%%%%%%%%%%%%%%% 

\clearpage
\onecolumn

\begin{table}

\centering
\caption{Log of Nov 2010 and May 2011 runs}
\label{tab:log}

\scalebox{0.8}{
\begin{tabular}{|c|c|c|c|c|c|}
\hline\hline 

Date & Mask & FWHM('') & frame(airmass) & t\_exposure(t\_ramp) & N$_{exposure}$ \\

\hline
Nov, 1-2 & ECDF-S 1&  & dark & 1s(1s), 1s(1s), 3s(1s), 300s(5s) &  10, 5, 5, 10 \\  %(5s)
Nov, 1-2& ECDF-S 1 & & lamp & 1s(1s) & 10 \\
Nov, 1-2 &ECDF-S 1 & & flat & 3s(1s) & 10 \\
Nov, 1-2 &ECDF-S 1 & 0.8 & NIR telluric standard & 1s(1s) & 20\\ %(5s)
Nov, 1-2 &ECDF-S 1 & 0.8 & science(1.13) & 300s(5s) & 58 \\
\hline
Nov, 1-2 & ECDF-S 2&  & dark &1s(1s), 300(5s) & 10, 10 \\ %(5s)
Nov, 1-2 &ECDF-S 2 &  & lamp & 1s(1s) & 10 \\ %(5s)
Nov, 1-2 &ECDF-S 2 &  & flat & 1s(1s) & 10 \\ %(5s)
Nov, 1-2 &ECDF-S 2 & 0.8 & NIR telluric standard & 1s(1s) & 25\\ %(5s)
Nov, 1-2 &ECDF-S 2 & 0.8 & science(1.15) & 300s(5s) & 56\\
\hline
May 19-20 & SDSS &  & dark &1s(1s), 3s(1s), 300s(5s) & 10, 10, 10 \\
May 19-20 & SDSS &  & lamp &1s(1s) & 5\\
May 19-20 & SDSS & & flat &3s(1s) & 5 \\
May 19-20 & SDSS & 1.0 & NIR telluric standard & 1s(1s) & 10 \\
May 19-20 & SDSS & 1.0 & science(1.2) &300s(5s) &38 \\
\hline
May 19-20 & EHDF-S &  & dark & 1s(1s), 3s(1s), 300s(5s) & 10, 10, 10\\
May 19-20 & EHDF-S & & lamp & 1s(1s) & 5\\
May 19-20 & EHDF-S &  & flat & 3s(1s) & 5\\
May 19-20 & EHDF-S & 1.0 & NIR telluric standard & 1s(1s) & 10 \\
May 19-20 & EHDF-S & 1.0 & science(1.3) &300s(5s) & 32\\
\hline

\end{tabular}
}
\tablefoot{The observations were carried out using the MMIRS $HK$ grism and the $H+K$ filter over 4 nights in three fields and four MOS masks.  Column 1: Observation date. Column 2: Mask name. For a few sources of ECDF-S 1 and ECDF-S 2 masks we used a narrower slit of 0.5''; a 0.7'' wide slit was used for the majority of the sources. Column 3: FWHM estimated from the continua of standard stars. The seeing was 0.8''(1'') in Nov(May) run. Column 4: type of calibration and science exposures performed for each mask (airmass is average). Column 5: Individual frame exposures; the ramp time is also specified for each type of frame in parentheses (Appendix A). %(\S \ref{sec:appendix})
Column 6: Number of each type of exposure. The total exposure time of the science frames was 4.8 (4.7) hours for ECDF-S mask1 (mask2), 3.2 (2.7) hours for SDSS (EHDF-S) mask.
}

\end{table}

\begin{table}

\centering
\caption{Sources detected in MMIRS survey}
\label{tab:det}

\scalebox{0.8}{
\begin{tabular}{|c|c|c|c|c|}

\hline\hline
objects & \#targeted & \#detected & \#emission line detected & \#continuum detected\\

%SFGs & & & &  \\
% &  &  &  Star  Forming  Galaxies &  \\
\hline
LAE & 27 & 4 & 4 &1   \\
\hline
DB &13 & 3 &3 & 0 \\
\hline
BX &29 & 6&6 & 0 \\
\hline
LBG &1 &1 &1  &0 \\
\hline
BzK &6 & 1& 1 &1 \\
\hline
%\hline
%LAEs & & & &  \\
%&  &  & Line Emitting Galaxies  &  \\
%\hline
\hline
AGN & 11&2 &2  &2 \\
%\hline
\hline
low-z & 5 &5 &0 & 3\\
\hline
slit stars & 3& 2 &0 &2 \\
\hline

\end{tabular}
}
\tablefoot{In this table we show the number of objects detected versus the ones targeted for each category of objects in the masks. The object name corresponds to the original category of the sources, as explained in \S \ref{sec:obs}. `low-z' are galaxies, whose BzK colors predict them to be at $z<1.4$. They appeared bright in the observed $H+K$ band and showed continua in MMIRS spectra. Together with slit stars these are useful to test the sensitivity and throughput of MMIRS.}

\end{table}

\begin{table}

\centering
\caption{List of sources showing emission lines in each mask}
\label{tab:list}

\scalebox{0.8}{
\begin{tabular}{|c|c|c|c|c|c|c|}
\hline\hline

%\tablehead{\colhead{\bf{name}}  &   \colhead{\bf{$\alpha$}}    &  \colhead{\bf{$\delta$}}        &    \colhead{\bf{$R_{AB}$}}  &\colhead{\bf{$H_{AB}$}}  &\colhead{\bf{$K_{AB}$}}  &\colhead {\bf{comments}} }

name & {\bf{$\alpha$}}    & {\bf{$\delta$}}        & {\bf{$R_{AB}$}}  & {\bf{$H_{AB}$}}  & {\bf{$K_{AB}$}}  & literature $z$\\

\hline
ECDF-S 1 & & & & & & \\
\hline
AGN5$^{b}$ & 3:32:31.46 &  -27:46:23.20 & 23.32 & 21.18 & 20.90 & $z=2.221$
 \\

BzK25$^{d}$ & 3:32:50.83 &   -27:48:04.68 & - & 23.49 & 23.05 &
%H_AB=23.49, K_AB=23.05  BzK, without Kvega<20
- \\ 

BX1$^{a}$ & 3:32:27.10  &  -27:46:43.25 & 24.65 & - &-  &  $z=2.223$ \\

BX10$^{a}$ & 3:32:33.78 &  -27:48:14.40 & 24.76 & & & $z=2.618$\\
%DBbright11 & 3:32:45.18 & -27:49:45.89 &26.53  &- & - & DBsli21$^{c}$\\

LAE27$^{e}$ & 3:32:52.68 &   -27:48:09.41 & 25.51 & 18.00 & 18.21 &  $z=3.085$\\
\hline
ECDF-S 2 & & & & & &\\
\hline
BX16$^{a}$ & 3:32:43.63 &   -27:43:47.71 & 24.03 & - & - & $z=2.317$\\

z3LAE2$^{e}$ & 3:32:26.94 &    -27:41:28.26 & 24.13 & - & - & $z=3.114$\\

DB8$^{c}$ & 3:32:36.14 &    -27:42:23.07 & - & - & 24.79 &  -\\
DB12$^{c}$ & 3:32:31.52 &   -27:43:51.38  & 24.51 & - & - & -\\
DB22$^{c}$ & 3:32:41.32 &  -27:45:49.13 &  26.43 & 24.88 & 25.58 & -\\
\hline
SDSS-1030 & & & & & & \\
\hline
AGN26 & 10:30:30.10 &   5:21:06.19 & 21.49 & 20.72 & 20.35&  -\\
BX4  & 10:30:10.25 &   5:22:15.77 & 25.23 & - & 22.79 &- \\
BX7 & 10:30:13.39 &   5:22:37.85 & 23.87 & 22.44 & 22.48 & - \\
BX14 & 10:30:18.86 &   5:21:40.87  & 24.18 & 22.95 & 22.33 & - \\
%BXwz30865(2.43) & 10:30:30.07 &   5:22:42.19 & 24.38 & 21.56 & 21.13 & BX24\\
z2LAE2 & 10:30:07.60 &   5:21:39.70 & 25.02 & - & 22.67 &  -\\
\hline
EHDF-S & & & & & & \\
\hline
z2LAE3 & 22:33:43.67 &  -60:39:44.56 &  24.25 &  - & - & -\\
LBG11 & 22:33:22.54 &   -60:38:49.63 & 24.55 & - & - &  $z=3.205$\\
\hline

\end{tabular}
}

\tablefoot{
%Under {\bf{name(z)}} column we indicate in parenthesis the redshift if a $z_{spec}$ was known for that source. 
Column 1: name of the source. Column 2 and 3: right ascension and declination. Column 4, 5, and 6:  source $R$, $H$, and $K$ magnitudes. Column 7: redshift if a $z_{spec}$ was known for that source. 
$^{a}$Lyman Break and BX galaxies are selected following selection criteria described in Ga2006 and Gu2010.
$^{b}$radio catalog object, observed through a 0.5" wide slit.
$^{c}$DB source from Hayes et al. (2010). 
$^{d}$%$K_{Vega}<20$ 
BzK source from Mark Dickinson and Jeyhan Kartaltepe (private communication).   
$^{e}z\simeq3.1$ LAEs from Gr2007.
The sources in SDSS-1030 and E-HDFS masks all belong to the MUSYC survey. 
The $R_{AB}$, $H_{AB}$, $K_{AB}$ magnitudes were obtained from the references above. The literature reshifts came either from the references above or from the public GOODS spectroscopic catalog (see text). Tab. \ref{tab:ELflux} lists our results.}

\end{table}

\addtocounter{table}{1}
\begin{table}

\caption{Galaxy metallicity}
\label{tab:MetEst}

\scalebox{0.8}{
\centering
\begin{tabular}{|c|c|c|c|c|c|c|c|c|c|}
\hline
\hline 

ID & $\frac{[OIII]5007}{H\beta}$ & $\rightarrow$  12+lg(O/H) & $\frac{[OIII]5007}{[NII]6584}$ & 
$\rightarrow$ 12+lg(O/H) & $\frac{[NII]6584}{H\alpha}$ & $\rightarrow$12+lg(O/H) & $\frac{[OIII]5007}{H\beta}/\frac{[NII]6584}{H\alpha}$ & $\rightarrow$ 12+lg(O/H) & 12+lg(O/H)$_{adopted}$ \\

(1)  & (2)  &  (3)   & (4) & (5)  & (6) & (7) & (8) & (9) &(10) \\ 
\hline
BX1 & $7.9\pm7.4$  &  [all]   & $7.9\pm13.9$ &  $\geq8.2$ & $0.4\pm0.7$ &  [all] &$20.3\pm40.7$&$\geq$8.1 &$\geq8.2$ \\ %Nagao 2006
  &   &     &  &   &  &  $\leq$8.9 & & & \\ %PP04
\hline
BX10 & $2.8\pm2.7$  & [all] & - & - & - & - & -& -& - \\
  &   & &  &  &  &-  &  & & \\
\hline
BX16 & $1.9\pm1.0$ & 8.6[8.4-8.8]  & $3.3\pm2.9$& 8.5[8.4-8.8]  & $0.16\pm0.14$ &8.6[7.9-8.9] & $12\pm12$ & $\geq8.2$  & $8.5^{+0.3}_{-0.1}$ \\
  &   &  $\leq$7.3   &  &   &  &  8.5[8.3-8.6] & & & \\ %PP04
\hline
BX4 & - &    - & - & - & $0.3\pm0.2$  & $\geq8.4$  & - & -  &
$\geq8.4$ \\
  &   &  -   &  &   &  &  8.6[8.4-8.7] & & & \\ %PP04
\hline
BX7 &  $\geq1.8$ & [7.2-8.6]   & $8.2\pm8.3$  & $\geq$8.1
&  $0.15\pm0.15$ & $\leq8.5$  & $\geq11$  & $\leq$8.3 &  $\leq8.5$ \\
  &   &     &  &   &  &  $\leq8.5$ & & & \\ %PP04
\hline
BX14 &  -  & - & $25.8\pm82.0$ & [all] & $0.05\pm0.14$ &
$\leq$8.7 & - & - & $\leq8.6$\\
  &   &  -  &  &   &  &  $\leq8.5$ & & & \\ %PP04
\hline
LBG11 & $2.8\pm2.4$ &$\leq$9.0 & - & - & -  & - &  - &  - & $\leq9.0$ \\
  &   &   &  &   &  &  - & & & \\ %PP04
\hline
LAE27 & $2.2\pm2.5$  & $\leq7.7$ & - & - & -  & - &  - &  - & - \\
  &   &  $\geq$8.1  &  &   &  &  - & & & \\ %PP04
\hline
z3LAE2 & $1.7\pm$0.8 & 7.1[7.0-7.2] & - & - & -  & - &  - &  - & $8.6^{+0.1}_{-0.1}$\\
  &   &  8.7[8.5-8.7]  &  &   &  &  - & & & \\ %PP04
\hline
z2LAE2 & $3.0\pm1.3$ &   [all] & - & - & -  & - &  - &  - & - \\
  &   &   &  &   &  &  - & & & \\ %PP04
\hline
DB8 &  $3.0\pm2.6$ &  [all] & - & - & -  & - &  - &  - & - \\
  &   &   &  &   &  &  - & & & \\ %PP04
\hline
DB12&  $2.9\pm2.3$ & [all] &$3.1\pm2.3$  & 8.5[8.4-8.9] & $0.22\pm0.15$ & 8.7[8.3-9.0]& $13.4\pm14.4$ & $\geq8.4$& $8.6^{+0.3}_{-0.3}$ \\ &   &    &  &   &  &  8.5[8.2-8.7] & & & \\ %PP04
\hline
DB22 & $\geq1.7$ & [7.1-8.7] & - & - & -  & - &  - &  - & - \\
  &   &   &  &   &  &  - & & & \\ %PP04
\hline

\end{tabular}
}
\tablefoot{
Column 1: names of the sources as presented in Table \ref{tab:list}. Column 2, 4, and 6: measured [OIII]5007/H$\beta$, [OIII]5007/[NII]6584, and [NII]6584/H$\alpha$ emission line ratios. Column 3, 5, and 7: metallicity, parametrized by 12+log(O/H) implied by the ratios reported in columns 2, 4, and 6 respectively. 
%In the 1st column we list the . In the 2nd, 4th, 6th, and 8th columns we report  and in the 3rd, 5th, 7th, 9th columns the implied metallicity, parametrized by 12+log(O/H). 
We used Nagao et al. (2006) calibrations for
  [OIII]5007/H$\beta$, dust-corrected [OIII]5007/[NII]6584, and
  [NII]6584/H$\alpha$. We adopted the calibrations from Pettini\&Pagel
  (2004) to infer [NII]6584/H$\alpha$ (second row of 7th column) and O3N2
  ([OIII]/H$\beta$ / [NII]/H$\alpha$). Column 9: the value of 12+log(O/H) we adopted for our sources as a combination
  of the information coming from the more conclusive ratios. In the
  case of z3LAE2 we considered the higher branch of
  [OIII]5007/H$\beta$ calibration the most representative of its
  metallicity, because EW(H$\beta$) was measured to be at least 10 {\AA} \citep{HuCowie2009}.}
\end{table}

\begin{table}

\caption{Galaxy properties}
\label{everypropV}

\scalebox{1.0}{
\centering 
\begin{tabular}{|c|c|c|c|c|c|c|c|c|c|}

\hline\hline
ID &  $\frac{H\alpha}{H\beta}$ & $\frac{Ly\alpha}{H\alpha}$ & $\frac{Ly\alpha}{[OIII]5007}$ & $\Delta v$ (km/sec) & E(B-V)$_{\beta}$ & E(B-V)$_{\beta-\alpha}$ &  E(B-V)$_{SED}$ & log(M$_*$/M$_{\odot}$) & 12+log(O/H) \\
(1)  & (2)  &  (3)   & (4) & (5)  & (6) & (7) & (8) & (9) &(10) \\ 
\hline
BX1 & $2.6\pm2.5$ & $2.5\pm0.6$ & $0.8\pm0.1$ &  $125.1\pm108.1$  & $0.12\pm0.02$ & $\leq0.54$ &  0.32$^{+0.06}_{-0.23}$ & 9.25$^{+0.3}_{-0.2}$ & $\geq8.2$  \\
\hline
BX10 & - & $\geq0.6$ & $0.49\pm0.17$ & $749.0\pm205.8$  &  $0.12\pm0.02$ & - &  0.32$^{+0.06}_{-0.23}$ & 9.21$^{+0.30}_{-0.20}$ & - \\
\hline
BX16 & $1.7\pm0.6$ & - & - & $-897.8\pm272.3$  &  $0.35\pm0.02$   & $\leq0.01$ &   0.10$^{+0.10}_{-0.30}$ & $9.9^{+0.1}_{-0.5}$ & $8.5^{+0.3}_{-0.1}$ \\
\hline
BX4 & -  & - & -  & - & - & - & 0.35$^{+0.02}_{-0.04}$ & $9.0^{+0.1}_{-0.1}$  & $\geq8.4$ \\
\hline
BX7 &   - & - & - &  - & -  & -     & 0.35$^{+0.02}_{-0.04}$ & $9.6^{+0.1}_{-0.1}$  & $\leq8.5$  \\
\hline
BX14 &  $4.0\pm3.2$ & - & - & - & - & $0.3\pm0.5$ &  0.35$^{+0.02}_{-0.04}$ & $9.4^{+0.1}_{-0.1}$  & $\leq8.6$ \\
\hline
LBG11 & -  & - &  $0.37\pm0.14$ & $344.5\pm107.0$ & $0.07\pm0.14$ & - &   $\geq0.1$  & $\geq9.95$ & $\leq9.0$ \\
\hline
LAE27 &  - & - & $0.08\pm0.02$ & $167.8\pm105.3$ & - & - & $\leq0.1$ & $9.95^{+0.13}_{-0.17}$ & - \\
\hline
z3LAE2 &  - & - & $0.11\pm0.03$ & $221.8\pm90.0$ & - & - & 0.32$^{+0.06}_{-0.23}$ & $9.95^{+0.13}_{-0.17}$ & $8.6^{+0.1}_{-0.1}$\\
\hline
z2LAE2 & -  & - & -  & - & -   & -  & 0.09$^{+0.26}_{-0.09}$ & $9.1^{+0.3}_{-0.2}$  & - \\
\hline
DB8 &  - & - & - &  - & - & - & 0.47$^{-0.09}_{+0.26}$ & $9.76^{+0.1}_{-0.1}$ & - \\
\hline
DB12&  $4.4\pm3.3$ & - & - & -& - & $0.42\pm0.20$ &  0.26$^{+0.26}_{-0.09}$ & $9.54^{+0.1}_{-0.1}$ & $8.6^{+0.3}_{-0.3}$ \\
\hline
DB22& - & -  & - & -  & - & -  & 0.22$^{+0.26}_{-0.09}$ & $9.26^{+0.1}_{-0.1}$  & - \\
\hline
\end{tabular}
}

\tablefoot{
Column 1: names of the sources as presented in Table \ref{tab:list}. Column 2, 3, and 4: H$\alpha$/H$\beta$, Ly$\alpha$/H$\alpha$, and Ly$\alpha$/[OIII]5007 line flux ratios. Column 5: velocity offsets calculated using equation (5). Column 6, 7, and 8: dust-reddening estimations, obtained from fitting the rest-frame UV slope (E(B-V)$_\beta$), the Balmer decrement (E(B-V)$_{\beta-\alpha}$) in the 2nd column, and from SED fitting (E(B-V)$_{SED}$). 
If continuum is too low to estimate $\beta$ slope we do not report E(B-V)$_{\beta}$.    
BX16 spectrum provides F(H$\alpha$)/F(H$\beta) = 1.7 \pm 0.6$, i.e. F(H$\alpha$)/F(H$\beta) < 2.9$ at 2$\sigma$, which is consistent with no reddening (E(B-V)$_{\beta-\alpha}\leq0.01$).
Column 9: stellar masses in unit of solar metallicity (M$_{*}$/M$_{\odot}$), inferred from the SED fit. Column 10: metallicity values taken from the final column of Table \ref{tab:MetEst}. 
} 
\end{table}

\clearpage
\onecolumn
\longtab{4}{
\begin{table}

\caption{Emission line fluxes for all the sources}
\label{tab:ELflux}

\begin{tabular}{|ccccccc|}

\hline %\hline
Line & $\lambda_{central}^{obs-frame}$ & $z_{spec}$  & Flux & rms f$_{\lambda}$ (continuum) & EW$_{obs-frame}$ & L\\
& \AA & & erg sec$^{-1}$ cm$^{-2}$ & erg sec$^{-1}$ cm$^{-2}$ \AA$^{-1}$ & \AA  & erg sec$^{-1}$ \\
\hline
{\bf{AGN5}} &  & & &  & &\\
\hline
H$\beta$&15679.3$\pm$11.5  & -  & (2.2$\pm$3.0)E-17 & 32.1E-19 & $\geq$7 & (0.5$\pm$1.1)E+42 \\
$[OIII]$4959& 15988.5$\pm$12.9 & - & (8.9$\pm$1.6)E-17& 32.1E-19 & $\geq$30 & (3.3$\pm$0.6)E+42 \\
$[OIII]$5007&16143.6$\pm$4.6 & $2.2234\pm0.0009$ & (26.6$\pm$4.7)E-17$^f$ & 32.1E-19  & $\geq$80 & (9.9$\pm$1.8)E+42 \\
\hline
$[NII]$6548&   21100.8$\pm$9.7 & - &  (7.2$\pm$1.2)E-17  &      22.3E-19  & $\geq$30  & (2.7$\pm$0.5)E+42 \\
H$\alpha$&   21153.1$\pm$6.9 & $2.2223\pm0.0011$  & (22.5$\pm$5.9)E-17$^f$ &       22.3E-19  & $\geq$100 & (8.6$\pm$2.2)E+42  \\
$[NII]$6583& 21217.2$\pm$11.1 & - & (21.7$\pm$3.2)E-17$^f$  &       22.3E-19  & $\geq$95 & (8.3$\pm$1.2)E+42 \\
\hline
$[SII]$doublet &     21651.8$\pm$2.7 & - & (8.7$\pm$5.2)E-17  &    22.3E-19  & $\geq$40 & (3.3$\pm$2.0)E+42 \\
\hline
\hline
{\bf{BzK25}}&  & & &  & &\\
\hline
Ly$\alpha$(R) & 3830.3$\pm$0.5  &  -   &  (24.3$\pm$2.0)E-17 & 24.0E-19  & 100 &   (8.5$\pm$0.7)E+42     \\
\hline
H$\beta$ &       15303.0$\pm$6.0 & - & (7.2$\pm$4.3)E-17 & 20.0E-19  & $\geq$18 &   (2.5$\pm$1.5)E+42  \\
$[OIII]$4959 &  15605.1$\pm$5.2 & - &  (12.7$\pm$1.9)E-17 & 20.0E-19  & $\geq$32 &  (4.4$\pm$0.7)E+42   \\
$[OIII]$5007 &  15756.2$\pm$1.8 & $2.1461\pm0.0003$ &  (37.7$\pm$5.7)E-17 & 20.0E-19  & $\geq$94 &  (13.2$\pm$2.0)E+42  \\
\hline
$[NII]$6548 & 20603.5$\pm$?? &  - &  (2.8$\pm$1.3)E-17$^f$ & 11.6E-19  &  $\geq$12 &  (1.0$\pm$0.5)E+42    \\
H$\alpha$ &   20650.8$\pm$4.2 & $2.1458\pm0.0006$ &   (25.1$\pm$2.6)E-17 &11.6E-19  & $\geq$100 &   (8.8$\pm$0.9)E+42   \\
$[NII]$6583 & 20713.6$\pm$78.9 & - &   (8.3$\pm$4.0)E-17 & 11.6E-19  & $\geq$36  & (2.9$\pm$1.4)E+42   \\
\hline
\hline
{\bf{AGN26}} & & & &  & & \\
\hline
H$\beta$ &  14669.3$\pm$14.7 & - & (22.7$\pm$8.5)E-17& 27.1E-19  &   130   & (6.8$\pm$2.5)E+42  \\
$[OIII]$4959 &  14958.9$\pm$12.6 & - &  (18.0$\pm$2.5)E-17 & 27.1E-19  & 100 &  (5.3$\pm$0.7)E+42   \\
$[OIII]$5007 & 15103.7$\pm$5.2 & $2.0158\pm0.0010$&  (53.5$\pm$7.5)E-17 & 27.1E-19  & 320 &  (16.0$\pm$2.2)E+42  \\
\hline
$[NII]$6548 &    19739.8$\pm$36.3 & - &    (20.5$\pm$3.6)E-17    & 24.0E-19  & 85 & (6.1$\pm$1.0)E+42  \\
H$\alpha$ & 19784.7$\pm$24.1 & $2.0138\pm0.0037$ &  (44.7$\pm$4.3)E-17 &   24.0E-19  & 180 &  (13.4$\pm$1.3)E+42  \\
$[NII]$6583 &    19845.0$\pm$46.4 & - &   (48.6$\pm$4.5)E-17      & 24.0E-19  & 200 & (14.5$\pm$1.3)E+42  \\
\hline
$[SII]$6718 &  20252.0$\pm$11.6 & - & (5.3$\pm$3.6)E-17$^f$  & 24.0E-19  & 20 &    (1.6$\pm$1.1)E+42  \\
$[SII]$6730 &  20288.1$\pm$17.5  & - &         (16.1$\pm$7.3)E-17$^f$   & 24.0E-19  & 65 &  (4.8$\pm$2.2)E+42  \\ 
\hline
\hline
{\bf{z2LAE3}} & & & &  & &\\
\hline
H$\beta$ &       14967.6$\pm$12.4 & - & (10.2$\pm$6.7)E-17    & 72.0E-19  &  $\geq$7  &  (3.3$\pm$2.2)E+42    \\ 
$[OIII]$4959 &  15263.1$\pm$6.5 & - &  (44.7$\pm$2.7)E-17$^f$    & 72.0E-19  & $\geq$30  &    (14.4$\pm$0.9)E+42   \\ 
$[OIII]$5007 &  15410.8$\pm$1.9 & $2.0771\pm0.0004$ &   (133.1$\pm$7.9)E-17    & 72.0E-19  &  $\geq$90 &  (43.0$\pm$2.5)E+42  \\ 
\hline
$[NII]$6548 & 20157.2$\pm$?? & -  &     (5.0$\pm$1.0)E-17    & 34.9E-19  &   $\geq$7 &  (1.6$\pm$0.3)E+42  \\ 
H$\alpha$ &   20203.1$\pm$4.9 & $2.0776\pm0.0007$ &   (58.6$\pm$4.1)E-17 &  34.9E-19  &  $\geq$80  &  (18.9$\pm$1.3)E+42  \\ 
$[NII]$6583 & 20265.9$\pm$7.8 & - &   (14.9$\pm$3.1)E-17$^f$     & 34.9E-19  & $\geq$20  &   (4.8$\pm$1.0)E+42  \\ 
\hline
\hline
\hline
\hline
&  & & &  & &\\
{\bf{BX1}}&  & & &  & &\\
\hline
Ly$\alpha$(R) & 3921.8$\pm$0.6  &  $2.2260\pm0.0005$   &  (10.1$\pm$0.8)E-17 & 4.9E-19  & 150 &  (3.9$\pm$0.3)E+42\\
\hline
H$\beta$&     15691.5$\pm$28.8  & - & (1.6$\pm$1.5)E-17  &  10.3E-19      &  $\geq$8  &  (0.6$\pm$0.6)E+42 \\
$[OIII]$4959 & 15986.7$\pm$5.7 & - &  (4.3$\pm$0.4)E-17 & 10.3E-19 & $\geq$20  & (1.6$\pm$0.2)E+42  \\
$[OIII]$5007 & 16149.8$\pm$1.9 & $2.2246\pm0.0004$ &   (12.7$\pm$1.2)E-17 & 10.3E-19 &    $\geq$60    &  (4.8$\pm$0.5)E+42   \\
\hline
$[NII]$6548 &    21110.3$\pm$47.8 & - &  (0.5$\pm$1.0)E-17  & 7.0E-19 & $\geq$3  &  (0.2$\pm$0.3)E+42  \\
H$\alpha$   &    21158.3$\pm$5.4 &  $2.2231\pm0.0008$ & (4.1$\pm$0.9)E-17 &7.0E-19 & $\geq$28 &   (1.6$\pm$0.3)E+42   \\
$[NII]$6583 &    21224.1$\pm$?? & - & (1.6$\pm$2.8)E-17$^f$   & 7.0E-19 &   $\geq$11   &    (0.6$\pm$2.8)E+42    \\
\hline
\hline
{\bf{BX10}}&  & & &  & &\\
\hline
Ly$\alpha$(R) & 4403.4$\pm$0.3  &  $2.622\pm0.0002$   &  (9.7$\pm$0.5)E-17 & 7.3E-19  & 103 &   (5.5$\pm$0.3)E+42     \\
\hline
H$\beta$&      17575.0$\pm$19.9  & - & (7.0$\pm$6.3)E-17 &   44.6E-19  & $\geq$8  &   (4.0$\pm$0.4)E+42    \\
$[OIII]$4959& 17922.0$\pm$20.6 & - &(6.5$\pm$2.4)E-17  &   44.6E-19  &  $\geq$7 &  (3.7$\pm$1.4)E+42   \\
$[OIII]$5007&  18095.5$\pm$11.6 & $2.6132\pm0.0026$ & (19.3$\pm$7.0)E-17$^f$  &  44.6E-19  & $\geq$22  &  (11.0$\pm$4.0)E+42   \\
\hline
H$\alpha$&  6562.82$\times$3.614 & - &$\leq$17.5E-17  & 35.3E-19  &   - &  $\leq$9.9E+42   \\
\hline
\hline
{\bf{BX4}} & & & &  & &\\
\hline
$[OIII]$5007 &  5007.0$\times$3.505 & - &  ($\leq$10.0)E-17$^f$ &18.9E-19  &  - & $\leq$5.0E+42  \\
\hline
$[NII]$6548 & 22948.4$\pm$?? & - &   (1.9$\pm$1.0)E-17 & 20.1E-19  & $\geq$5  &   (0.9$\pm$0.5)E+42  \\
H$\alpha$ &    23000.9$\pm$8.7 & $2.5038\pm0.0013$ & (16.6$\pm$3.3)E-17 &20.1E-19  &   $\geq$40  &  (8.2$\pm$1.6)E+42 \\
$[NII]$6583 &  23071.0$\pm$10.6 & -&   (5.8$\pm$3.0)E-17 & 20.1E-19  &     $\geq$14    &  (2.9$\pm$1.5)E+42  \\
\hline
\hline
{\bf{BX7}} & & & &  & &\\
\hline
H$\beta$ &  4863.0$\times$3.418 &  - &  $\leq$7.5E-17 & 15.4E-19  &   -    & $\leq$3.4E+42  \\
$[OIII]$4959 &  16949.7$\pm$10.0 & - &  (4.4$\pm$0.7)E-17 & 15.4E-19  & $\geq$13   &   (2.0$\pm$0.3)E+42  \\
$[OIII]$5007 &  17113.8$\pm$3.6 & $2.4171\pm0.0007$ &  (13.0$\pm$2.4)E-17$^f$ & 15.4E-19  &   $\geq$40  &  (6.0$\pm$1.0)E+42   \\
\hline
$[NII]$6548 & 22362.9$\pm$82.1 & - & (0.5$\pm$0.5)E-17    & 10.7E-19  &   $\geq$1  &  (0.2$\pm$0.2)E+42  \\
H$\alpha$ &   22413.8$\pm$4.0 & $2.4143\pm0.0006$ & (10.5$\pm$1.6)E-17    & 10.7E-19  &  $\geq$50   &  (4.8$\pm$0.7)E+42 \\
$[NII]$6583  & 22483.4$\pm$32.7 & -&  (1.6$\pm$1.6)E-17$^f$    & 10.7E-19  & $\geq$7   &  (0.7$\pm$0.7)E+42   \\
\hline

\end{tabular}

\tablefoot{$^f$ means sky line residuals can affect the flux estimation. (R) refers to the fact that the rest-frame UV spectrum was normalized to the galaxy $R$ band magnitude. Ly$\alpha$ emission line fluxes of `BX' galaxies are estimated from
GOODS public spectra (5.7
{\AA}/pixel, VIMOS-LR); of LAE27 are calculated from spectra obtained with VIMOS (2.57 {\AA}/pixel,
VIMOS-MR), of z3LAE2 and LBG11 with IMACS (1.3
{\AA}/pixel) by MUSYC. Optical emission lines are obtained from our 7
{\AA}/pixel spectra. To determine redshifts we assumed vacuum 
$\lambda_{Ly\alpha}=1215.67$ {\AA}, $\lambda_{[OIII]5007}=5008.239 $
{\AA}, $\lambda_{H\alpha}=6564.614$ {\AA}. %For DBslit12 I did not use
                                %clipping!
Column 1, 2, and 3: emission line names, their central wavelengths, and implied redshift. An `??' indicates a formally infinite error. Column 4: emission line integrated fluxes. Column 5: continuum flux densities. Column 6: obs-frame equivalent widths of the emission lines listed in the 1st column. In the case of lower limits, the equivalent width is estimated from the 2$\sigma$ upper limit of the continuum. The continuum $\sigma$ is defined as its root-mean-square.  Column 7: emission line luminosities. For a few of the DB sources we report photometric H$\alpha$ luminosity as $^{phot}$.  }

\end{table}

}

\longtab{4}{
\begin{table}

\caption{Continued: Emission line fluxes for all the sources}
%\label{tab:ELflux}

\begin{tabular}{|ccccccc|}

\hline %\hline
Line & $\lambda_{central}^{obs-frame}$ & $z_{spec}$  & Flux & rms f$_{\lambda}$ (continuum) & EW$_{obs-frame}$ & L\\
& \AA & & erg sec$^{-1}$ cm$^{-2}$ & erg sec$^{-1}$ cm$^{-2}$ \AA$^{-1}$ & \AA  & erg sec$^{-1}$ \\
\hline

{\bf{BX14}}& & & &  & &\\
\hline
H$\beta$ &   16373.0$\pm$9.6 &  - &   (3.3$\pm$2.6)E-17$^f$ & 20.8E-19  &   $\geq$15   &   (1.4$\pm$1.1)E+42  \\
$[OIII]$4959 &  16696.2$\pm$11.7 & - &  (5.2$\pm$1.0)E-17$^f$ & 20.8E-19  & $\geq$24   &  (2.2$\pm$0.4)E+42  \\
$[OIII]$5007 &  16857.8$\pm$3.9 &  $2.3660\pm0.0008$&   (15.6$\pm$3.0)E-17    & 20.8E-19  &  $\geq$70  &   (6.7$\pm$1.3)E+42   \\
\hline
$[NII]$6548 & 22049.7$\pm$?? & - & (0.2$\pm$0.6)E-17$^f$    &      14.0E-19  &   $\geq$0.7 &  (0.08$\pm$0.3)E+42   \\
H$\alpha$ & 22099.9$\pm$4.1 & $2.3665\pm0.0006$ & (13.3$\pm$2.3)E-17  &   14.0E-19  & $\geq$47   &  (5.7$\pm$1.0)E+42  \\
$[NII]$6583 & 22168.6$\pm$51.6 & - &    (0.6$\pm$1.9)E-17$^f$       & 14.0E-19  &  $\geq$2  & (0.3$\pm$0.8)E+42 \\
\hline
\hline
{\bf{BX16}}&  & & &  & &\\
\hline
Ly$\alpha$(R) & 4019.0$\pm$1.2  &  $2.3060\pm0.0010$   &  -(6.3$\pm$1.1)E-17 & 12.3E-19  & -45 &   -  \\
\hline
H$\beta$ &       16129.1$\pm$?? & - & (2.9$\pm$1.3)E-17 & 13.0E-19  &  $\geq$20    &  (2.4$\pm$0.7)E+42  \\
$[OIII]$4959 &  16447.5$\pm$15.0 & - &  (1.9$\pm$0.6)E-17$^f$ & 13.0E-19  & $\geq$6  & (0.8$\pm$0.2)E+42   \\
$[OIII]$5007 &  16606.7$\pm$5.7  & $2.3159\pm0.0011$ &  (5.6$\pm$1.7)E-17 & 13.0E-19  &  $\geq$40 &  (1.2$\pm$0.5)E+42    \\
\hline
$[NII]$6548 & 21718.1$\pm$?? & - &   (0.6$\pm$0.4)E-17$^f$ & 11.5E-19  &  $\geq$3  &   (0.3$\pm$0.2)E+42    \\
H$\alpha$ &    21767.5$\pm$4.3 & $2.3159\pm0.0007$ & (10.4$\pm$1.6)E-17 & 11.5E-19  &   $\geq$45 & (4.3$\pm$0.7)E+42   \\
$[NII]$6583 &  21835.2$\pm$13.4 & - &   (1.7$\pm$1.4)E-17$^f$ & 11.5E-19  &  $\geq$7  &    (0.7$\pm$0.6)E+42    \\
\hline
\hline
{\bf{LBG11}}  & & & &  & &\\
\hline
Ly$\alpha$(R) & 5112.4$\pm$0.6  &  $3.2054\pm0.0005$  &  (3.0$\pm$0.7)E-17 & 11.6E-19  & 34 &  (2.8$\pm$0.7)E+42   \\
\hline
H$\beta$ &       20432.6$\pm$14.4 & - & (2.9$\pm$2.4)E-17    & 8.2E-19  &  $\geq$18 &   (2.7$\pm$2.2)E+42  \\
$[OIII]$4959 &  20836.0$\pm$16.0 & - &  (2.6$\pm$0.8)E-17$^f$    & 8.2E-19  & $\geq$16 &  (2.4$\pm$0.7)E+42  \\
$[OIII]$5007 &  21037.7$\pm$5.0 & $3.2006\pm0.0010$ &  (8.0$\pm$2.3)E-17    & 8.2E-19  & $\geq$50 &  (7.5$\pm$2.1)E+42  \\
\hline
\hline
{\bf{LAE27}}&  & & &  & &\\
\hline
Ly$\alpha$(R) & 4966.7$\pm$0.3  &  $3.0856\pm0.0002$   &  (2.2$\pm$0.2)E-17 & 4.4E-19  & 105 &    (1.9$\pm$0.2)E+42   \\
\hline
H$\beta$ &       19860.8$\pm$48.8 & - & (12.0$\pm$13.3)E-17  & 32.5E-19  &   $\geq$19   &   (10.1$\pm$11.3)E+42   \\
$[OIII]$4959 &  20252.8$\pm$16.4 & - &  (8.2$\pm$2.6)E-17$^f$ & 32.5E-19  &  $\geq$14  & (7.0$\pm$2.2)E+42   \\
$[OIII]$5007 &  20448.9$\pm$5.0 & $3.0830\pm0.0001$ &  (25.8$\pm$7.7)E-17 & 32.5E-19  &   $\geq$41    &   (21.9$\pm$6.5)E+42   \\
\hline
\hline
{\bf{z3LAE2}}&  & & &  & &\\
\hline
Ly$\alpha$(R) & 5002.4$\pm$0.5  &  $3.1149\pm0.0004$   &  (2.3$\pm$0.5)E-17 & 9.9E-19  & 74 &    (2.0$\pm$0.4)E+42   \\
\hline
H$\beta$ &       20000.5$\pm$11.4 & - & (12.3$\pm$5.0)E-17$^f$  & 29.6E-19  & $\geq$20   &   (10.7$\pm$6.8)E+42   \\
$[OIII]$4959 &  20395.3$\pm$11.1 & - &  (7.0$\pm$1.5)E-17$^f$ & 29.6E-19  & $\geq$12 &  (6.1$\pm$1.3)E+42    \\
$[OIII]$5007 &  20592.7$\pm$3.1 & $3.1118\pm0.0006$ &  (20.9$\pm$4.4)E-17  &29.6E-19  &  $\geq$35 &  (18.2$\pm$3.8)E+42    \\
\hline
\hline
{\bf{z2LAE2}} & & & &  & &\\
\hline
H$\beta$ &        14849.6$\pm$5.7 &  - &   (11.7$\pm$4.9)E-17$^f$ & 20.0E-19  &  $\geq$30   & (3.7$\pm$1.6)E+42  \\
$[OIII]$4959 &  15142.8$\pm$10.3 & - &  (11.4$\pm$1.5)E-17 & 20.0E-19  &   $\geq$30 & (3.6$\pm$0.5)E+42  \\
$[OIII]$5007 &  15289.3$\pm$4.4 & $2.0528\pm0.0009$ &  (34.1$\pm$4.6)E-17 &20.0E-19  &   $\geq$90 & (10.8$\pm$1.5)E+42 \\
\hline
H$\alpha$ &    6562.82$\times$3.054 & - & $\leq$15.0E-17$^f$ & 28.0E-19  &  -  & $\leq$4.8E+42   \\
\hline
\hline
{\bf{DB8}} &  & & &  & &\\
\hline
Ly$\alpha$ &    &    &   &  &  & $\leq$2.8E+41$^{phot}$   \\
\hline
H$\beta$ &       15952.8$\pm$30.9 & - & (4.3$\pm$3.4)E-17 & 15.0E-19  &   $\geq$14   &   (1.8$\pm$1.4)E+42    \\

$[OIII]$4959 &   16288.1$\pm$28.2 & - &  (4.1$\pm$0.9)E-17$^f$ & 15.0E-19  & $\geq$15  & (1.7$\pm$0.4)E+42   \\

$[OIII]$5007 &  16445.8$\pm$8.0 & $2.2837\pm0.0015$ &  (12.5$\pm$2.8)E-17 & 15.0E-19  & $\geq$45  &  (5.1$\pm$1.1)E+42   \\
\hline
H$\alpha$ &       6562.82$\times$3.285 & - & $\leq$5.0E-17 & 9.5E-19  & - & 2.3E+41$^{phot}$ \\
\hline
\hline
{\bf{DB12}}&  & & &  & &\\
\hline
Ly$\alpha$ &    &    &   &  &  & 1.6E+42$^{phot}$   \\
\hline
H$\beta$ &       15518.0$\pm$10.1 & - & (4.1$\pm$3.0)E-17 & 16.0E-19  &  $\geq$12  &  (1.5$\pm$1.1)E+42    \\
$[OIII]$4959 &  15824.3$\pm$29.3 & - &  (4.0$\pm$1.4)E-17$^f$ & 16.0E-19  &   $\geq$12  &   (1.5$\pm$0.5)E+42 \\
$[OIII]$5007 &  15977.5$\pm$7.0 & $2.1902\pm0.0014$ &  (12.0$\pm$4.0)E-17 & 16.0E-19  & $\geq$27&  (4.4$\pm$1.5)E+42  \\
\hline
$[NII]$6548 & 20907.8$\pm$10.4 & - &    (1.3$\pm$0.9)E-17 & 18.3E-19  &  $\geq$3  &  (0.5$\pm$0.3)E+42 \\
H$\alpha$ &   20955.4$\pm$8.0 & $2.1922\pm0.0012$ &  (18.1$\pm$3.2)E-17 & 18.3E-19  &  $\geq$50  & (6.7$\pm$1.2)E+42  \\
$[NII]$6583 & 21020.5$\pm$33.0 & - &    (3.9$\pm$2.6)E-17 & 18.3E-19  & $\geq$10   & (1.4$\pm$0.9)E+42  \\
\hline
\hline
{\bf{DB22}} &  & & &  & &\\
\hline
Ly$\alpha$ &    &    &   &  &  & $\leq$2.8E+41$^{phot}$   \\
\hline
H$\beta$ &       4863.0$\times$3.212 & - & $\leq$12.5E-17 & 22.8E-19  &  -  &  $\leq$4.9E+42  \\
$[OIII]$4959 &  15930.3$\pm$34.4 & - &  (6.9$\pm$1.7)E-17 & 22.8E-19  &   $\geq$14  &  (2.7$\pm$0.7)E+42  \\
$[OIII]$5007 &  16084.5$\pm$11.8  & $2.2116\pm0.0024$ &  (20.7$\pm$5.7)E-17 & 22.8E-19  &  $\geq$40 & (8.0$\pm$2.2)E+42 \\
\hline
H$\alpha$ &    6562.82$\times$3.212 & - & $\leq$6.0E-17$^f$ & 13.1E-19  & - & 3.9E+41$^{phot}$ \\
\hline

\end{tabular}

\end{table}

}

\appendix

\section{MMIRS data reduction and performance}
\label{sec:appendix}

\subsection{Reduction procedure}
\label{sec:reduction}

The raw mask frames, output of MMIRS, are multi-extension files. The 3nd to Nth extensions are characterized by individual exposure times equal to t\_ramp up to t\_total$-$t\_ramp, where t\_ramp is the ramp-up time and t\_total is the total exposure time of the frame. The read-out is done using a ramp-up procedure, optimized to avoid saturation. It is made every t\_ramp seconds, t\_total/t\_ramp times. The 1st extension in the multi-extension MMIRS frame contains the total exposure time, while the 2nd contains the zero-level counts.  

Dark frames are obtained in the morning for every kind of exposure taken during the previous night and in series of 5 (ramped-up multi-extension darks should be subtracted from the ramped-up science frames). We average combined multi-extension dark frames, using IRAF $mscred.combine$ task.
 
The following step of our reduction pipeline is collapsing all of the information enclosed in the multi-extension files. We used $mmfixen$ code. The information contained in the N extensions are combined, saturated pixels are filled and cosmic rays are removed. The output is a new multi-extension file, in which the 1st extension is the science frame normalized to counts/sec, the 2nd, 3rd and 4th contain statistical information about the image and the fitting process.

We ran $mmfixen$ code for all the ramped-up science frames and then we extracted their 1st extensions. In the case of t\_total$<$t\_ramp we just subtracted the zero-level counts (2nd extension) to the 1st extension. It happened for the majority of flat frames and for standard stars. We then normalized flat frames, slit by slit, using IRAF $twodspec.longslit.response$ task and we divided science and standard frames by the normalized flat. 
 
COSMOS software was used to generate the wavelength map needed for wavelength calibration. 
We first obtained a rough wavelength map using the calibration lamp. The spectra of an Argon \footnote{MMIRS/MMIRS+Pipeline/Wavelength+Calibration/Argon+Lines} lamp are obtained through each slit of a mask. 
Lamp emission line pixels were related to known wavelengths  
using COSMOS  $align$-$mask$ and $adjust$-$offset$ codes. An improved wavelength solution and spectral map were calculated using a few isolated sky lines in science frames. \footnote{MMIRS/MMIRS+Pipeline/Wavelength+Calibration/OH+Sky+Lines} 
We then ran COSMOS $adjust$-$map$ to obtain the final solution, which contains the best wavelength map for all the mask slits together and it is satisfied by the entire series of sky lines from $\sim$14000 to $\sim$22000 {\AA}.  

We applied an $ABBA$ procedure to get rid of prominent sky lines. 
In our May, 2011 run, sky line variations and along individual lines were stronger. Therefore, we got a better sky subtraction in Nov frames.
We had the case of the first night of May run in which SDSS mask frames were not ramped-up. We just dark-subtracted and flat-divided them and directly applied this ABBA procedure. We then removed cosmic rays using IRAF $xdimsum.xzap$ task to ABBA frames to eliminate cosmic rays, coming from A, and to the inverse of ABBA to eliminate cosmic rays, coming from B. We checked the alignment of the ABBA frames, looking at the continuum spectra of alignment stars (also slit stars in the case of May run) and combined all the aligned ABBA frames of the same mask. 

We finally applied COSMOS $extract$-$2dspec$ code to wavelength calibrate the completely reduced frame and divide it in individual-slit 2D spectra. 

\subsection{Flux calibration}
\label{sec:fluxcal}

We used IRAF $twodspec.apextract.apall$ task to extract 1D spectra. For standard stars we located the extraction aperture in a region of their continuum not affected by atmospheric absorption. For science spectra we chose the pixel position of the few prominent emission lines.

The three standard stars chosen at the observatory are telluric stars. Therefore they allowed to perform the correction for telluric features of science frames and also their flux calibration. Spectra coming from the same star, but obtained through different slits along a mask,  cover different wavelength regions, depending on the position of the slit along the dispersion axis. They present comparable shapes in the common wavelength regions, but their relative intensity depends on the mask orientation and on the sky conditions, that can change with time. These variations are reflected in the science frames as well. So we average-combined all the 1D spectra of the same telluric stars of the run, taken at the same airmass of the science frames.

\begin{figure*}[h!]
\centering
\includegraphics[width=120mm]{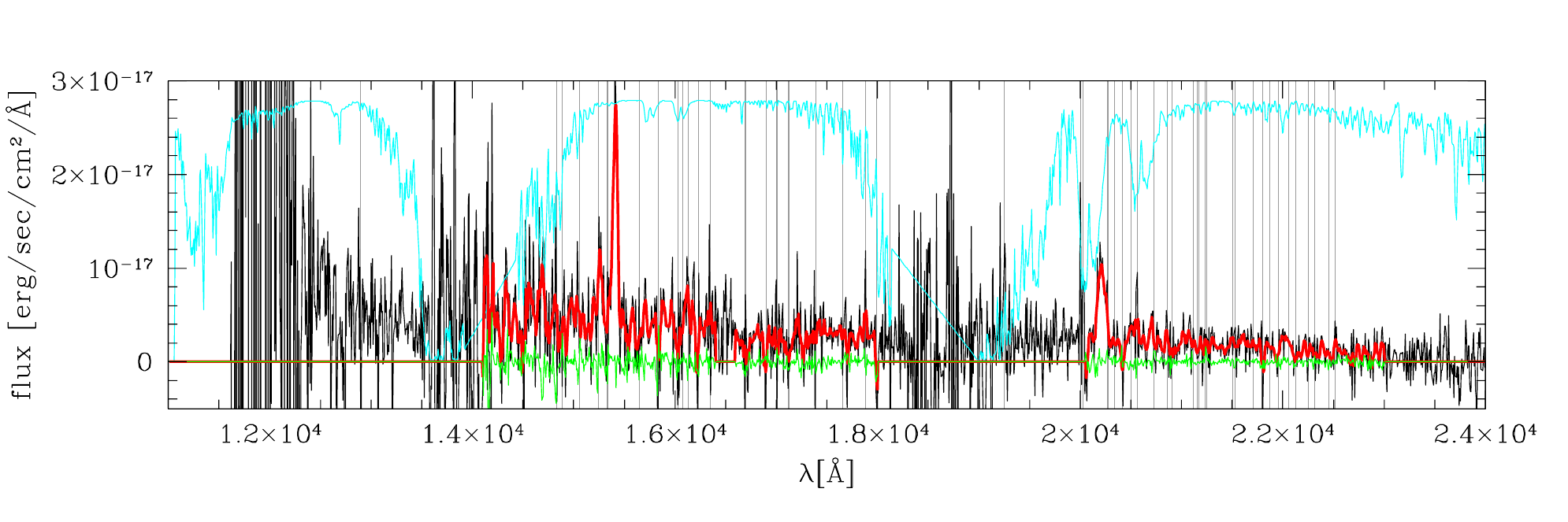}
\caption{1D spectrum of z2LAE3 along the entire MMIRS coverage. The black (red) curve represents the raw (smoothed) spectrum. Grey vertical lines indicate the wavelength of the dominant sky emission lines. Cyan curve represents the atmospheric transmission at Las Campanas observatory. At $\lambda \sim$ 14000 and $\sim$ 19000 {\AA} the transmission is very low. The green curve represents the background spectrum extracted far from the center of the slit where the source is positioned. Sky line residuals produce higher uncertainty. Telluric correction fixes the continuum shape but increases the uncertainty. Wavelength regions of high noise are not shown in the smoothed spectrum.}
\label{h2LAEslit2Nofit}
\end{figure*}

We extracted 1D spectra of the objects in the center of the slits as well as the ones of the background either on the upper or lower side of the slits to estimate background fluctuations, after sky emission line removal. 
In Fig. \ref{h2LAEslit2Nofit} we show z2LAE3 raw(smoothed) spectrum in black(red) and the background spectrum in green. This background spectrum is a proxy of the noise spectrum of this object.

\section{MMIRS 2D spectra and performance}

\begin{figure*} 
\centering
\includegraphics[width=150mm]{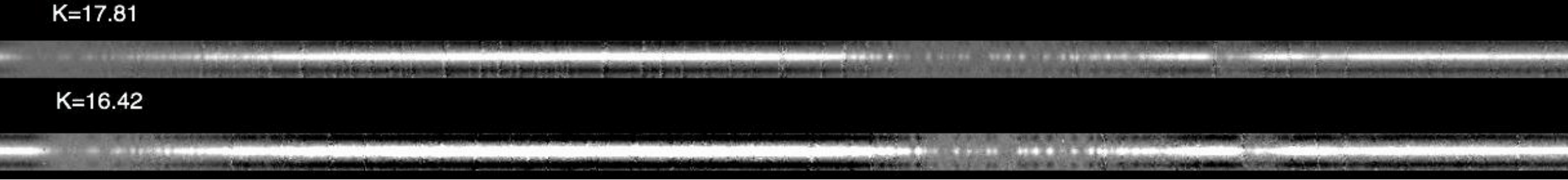}
\caption{2D spectra of two slit stars in the SDSS and EHDF-S masks with $K_{AB}=17.81$ (up) and $K_{AB}=16.42$ (low), respectively. }
\label{stars}
\end{figure*}

\begin{figure*} 
\centering
\includegraphics[width=150mm]{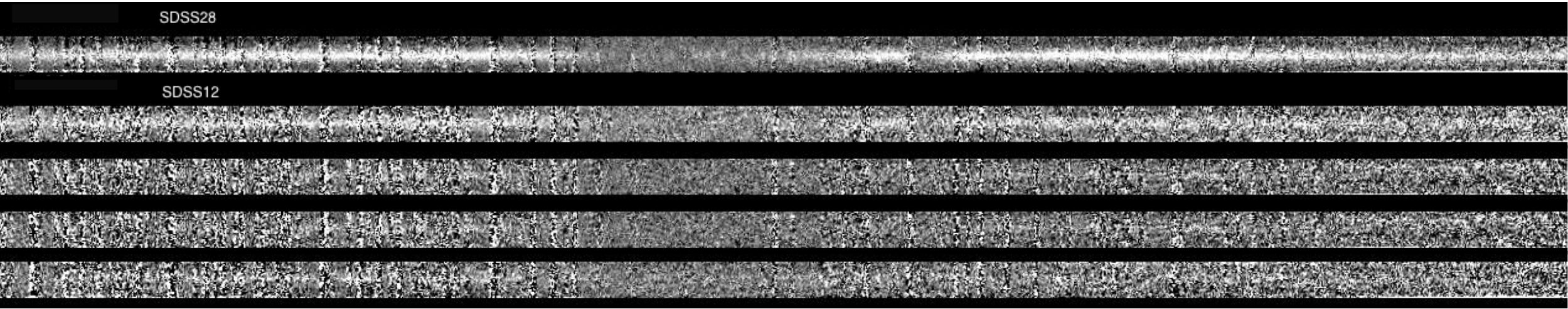}
\caption{2D spectra of the 'low-z', low-redshift galaxies obtained with SDSS mask. The spectra are aligned in wavelength. The top panel shows SDSS28 with the highest $K_{AB}$ band mag of 17.65. 
These sources spanned $K_{AB}$ band magnitudes from 17.65 (SDSS28), 19.39 (SDSS12) to 20.46. 
 In the center of the stripes ($\sim$19000 {\AA}), the decrease in flux indicates the wavelength region of strong atmospheric absorption. We can also see regions of high Poisson noise due to sky emission lines. }
\label{nonBzK}
\end{figure*}

As we could see in Table \ref{tab:det}, $z<1.4$ `low-z' galaxies and slit stars were located in MMIRS masks as a test of its performance. 
Fig. \ref{stars} shows 2D spectra of two slit stars in SDSS (upper) and EHDF-S (lower) masks. A star of $K_{AB}=16.42$ starts to show saturation in an MMIRS spectrum, while the continuum of a $K_{AB}=17.81$ star is clearly detected. Regions of strong atmospheric absorption are clearly seen, at $\sim$14000 {\AA} (in the left side of the spectra), at $\sim$19000 {\AA} (in the center-right side of the spactra). In these specific cases the slits were located at the same position along the dispersion axis of MMIRS mask.
In Fig. \ref{nonBzK} we present the 2D spectra of 5 ’low-z’ galaxies obtained with SDSS mask. The top panel shows the source with the brightest $K_{AB}$ band mag of 17.65. The continuum appears well detected for sources with $K_{AB}\leq19.39$.

%END
\end{document}